\documentclass[twocolumn, tighten]{aastex631}

\newcommand{\FIRE}{{\fontfamily{qcr}\selectfont
FIRE}}

\newcommand{\FIREtwo}{{\fontfamily{qcr}\selectfont
FIRE-2 }}
\newcommand{\FIREone}{{\fontfamily{qcr}\selectfont
FIRE-1 }}

\newcommand{\SKIRT}{{\fontfamily{qcr}\selectfont
SKIRT }}

\newcommand{\kmps}{$\mathrm{km\,s^{-1}}$}
\newcommand{\microm}{$\mathrm{\upmu m}$}
\newcommand{\msun}{M$_{\odot}$}

\usepackage{placeins} 
\usepackage{upgreek}
\usepackage{lineno}
\usepackage{comment} 
\usepackage[normalem]{ulem}
\graphicspath{{./}{figures/}}

\received{\today}

\accepted{TBD}

\hypersetup{linkcolor=blue,citecolor=blue,filecolor=blue,urlcolor=blue}

\shorttitle{Exploring ultra-compact early galaxies in cosmological FIRE simulations}
\shortauthors{Roy et al.}

\begin{document}

\title{Little Red Dots on FIRE: \\ Exploring the formation and observational signatures of ultra-compact early galaxies}

\correspondingauthor{Niranjan Chandra Roy}
\email{niranjan.roy@uconn.edu}

\author[0000-0002-0487-3090]{Niranjan Chandra Roy}
\affiliation{Department of Physics, University of Connecticut, 196 Auditorium Road, U-3046, Storrs, CT 06269-3046, USA}

\author[0000-0001-5769-4945]{Daniel Anglés-Alcázar}
\affiliation{Department of Physics, University of Connecticut, 196 Auditorium Road, U-3046, Storrs, CT 06269-3046, USA}

\author[0000-0001-8855-6107
]{Rachel K. Cochrane}
\affiliation{Jodrell Bank Centre for Astrophysics, University of Manchester, Oxford Road, Manchester M13 9PL, UK}

\author[0000-0003-0502-9235]{Alexander J. Richings}
\affiliation{Centre for Data Science, Artificial Intelligence and Modelling, University of Hull, Cottingham Road, Hull, HU6 7RX, UK}
\affiliation{E. A. Milne Centre for Astrophysics, University of Hull, Cottingham Road, Hull, HU6 7RX, UK}

\author[0000-0002-8390-6726]{Jonathan Mercedes-Feliz}
\affiliation{Department of Physics, University of Connecticut, 196 Auditorium Road, U-3046, Storrs, CT 06269-3046, USA}

\author[0000-0003-4073-3236]{Christopher C. Hayward}
\affiliation{J.P. Morgan Securities, 390 Madison Avenue, New York, NY, 10017, USA
}

\author[0000-0003-0502-9235]{Claude-André Faucher-Gigu\`{e}re}
\affiliation{CIERA and Department of Physics and Astronomy, Northwestern University, 1800 Sherman Avenue, 8th Floor CIERA, Evanston, IL 60201, USA}

\author[0000-0002-6149-8178]{Erini Lambrides}
\altaffiliation{NPP Fellow}
\affiliation{NASA-Goddard Space Flight Center, Code 662, Greenbelt, MD 20771, USA}

\author[0000-0002-1109-1919]{Robert Feldmann}
\affiliation{Department of Astrophysics, University of Zurich, Winterthurerstrasse 190, Zurich CH-8057, Switzerland}

\author[0000-0003-4597-6739]{Boon Kiat Oh}
\affiliation{Department of Physics, University of Connecticut, 196 Auditorium Road, U-3046, Storrs, CT 06269-3046, USA}
\affiliation{School of Physics, Korea Institute for Advanced Study, 85 Hoegiro, Dongdaemun-gu, Seoul 02455, Republic of Korea}

\author[0000-0003-0502-9235]{Andrew Marszewski}
\affiliation{CIERA and Department of Physics and Astronomy, Northwestern University, 1800 Sherman Avenue, 8th Floor CIERA, Evanston, IL 60201, USA}

\author[0000-0003-0502-9235]{Guochao Sun}
\affiliation{CIERA and Department of Physics and Astronomy, Northwestern University, 1800 Sherman Avenue, 8th Floor CIERA, Evanston, IL 60201, USA}

\author[0000-0001-8047-8351]{Kelcey Davis}
\altaffiliation{NSF Graduate Research Fellow}
\affiliation{Department of Physics, University of Connecticut, 196 Auditorium Road, U-3046, Storrs, CT 06269-3046, USA}
\affil{Los Alamos National Laboratory, Los Alamos, NM 87545, USA}

\author[0000-0002-6149-8178]{Jed McKinney}
\altaffiliation{NASA Hubble Fellow}
\affiliation{Department of Astronomy, The University of Texas at Austin, Austin, TX, USA}
\affiliation{Department of Physics, University of California Santa Barbara, Santa Barbara, CA 93106, USA}

\author[0000-0002-0930-6466]{Caitlin M. Casey}
\affiliation{Department of Physics, University of California Santa Barbara, Santa Barbara, CA 93106, USA}

\author[0000-0003-0699-6083]{Tanio Díaz-Santos}
\affiliation{Institute of Astrophysics, Foundation for Research and Technology-Hellas (FORTH), Heraklion, 70013, Greece}
\affiliation{School of Sciences, European University Cyprus, Diogenes Street, Engomi, 1516 Nicosia, Cyprus}

\author[0000-0001-5384-3616]{Madisyn Brooks}
\affiliation{Department of Physics, University of Connecticut, 196 Auditorium Road, U-3046, Storrs, CT 06269-3046, USA}

\author[0009-0005-2334-4513]{Grace Farrell}
\affiliation{Department of Physics, University of Connecticut, 196 Auditorium Road, U-3046, Storrs, CT 06269-3046, USA}
\affiliation{Department of Astronomy and Planetary Science, Northern Arizona University, Flagstaff, AZ 86011, USA}

\begin{abstract}

Little Red Dots (LRDs) are compact sources with broad Balmer lines, Balmer breaks, anomalous UV emission, rising red continuum, and uncertain origin. We use FIRE cosmological simulations, 3D dust radiative transfer, and synthetic emission-line data cubes to test whether ultra-compact early galaxies can reproduce LRD-like observables without invoking AGN. In progenitors of present-day group halos ($M_{\mathrm{halo}}>10^{13.5}$\,M$_{\odot}$), we identify transient phases at $z\approx 4$--8 lasting $\sim$150--400\,Myr in which strong dissipative inflows build massive ($M_{\star} \sim 10^{8.5}$--$10^{10.5}$\,M$_{\odot}$), UV-bright ($-23\lesssim\,M_{\rm UV}\lesssim\,-20$), ultra-compact (R$_{\mathrm{eff}}< 300$\,pc) stellar cores with extreme circular velocity ($V_{\rm circ}>500$\,km\,s$^{-1}$) and consistent with several LRD properties: strong Balmer breaks (F$_{\upnu}$(4200\r{A})/F$_{\upnu}$(3500\r{A}) $\sim$\,2); blue UV beta slopes ($\beta_{\mathrm{UV}} \approx -1.25$); dust masses; ALMA non-detections; and Balmer-line widths up to $\sim$1500\,km\,s$^{-1}$ broadened by galaxy-scale dynamics. However, stellar emission and host-galaxy kinematics alone do not reproduce the red rest-optical continuum, more extreme Balmer breaks ($\gtrsim$2.5) and line widths ($\gtrsim 2000$\,km s$^{-1}$), or the broad-Balmer/narrow-forbidden-line signature of broad-line AGN. The same ultra-compact conditions efficiently fuel central BHs, suggesting a hybrid stellar+AGN scenario in which compact stars explain the UV continuum, Balmer break, and intermediate line widths while AGN supply the red optical continuum and more extreme line properties. With halo masses $M_{\rm halo}\sim10^{11-12.5}$\,M$_\odot$ and comoving abundance $\sim2\times10^{-5}\,{\rm cMpc}^{-3}$ (for $\sim$20\% duty-cycle at $z\approx 4$--8), ultra-compact galaxies can contribute to the massive, bright LRD population.

\end{abstract}
\vspace{-10mm}
\keywords{Galaxy: evolution -- Galaxy: kinematics and dynamics -- Galaxies: high-redshift -- Galaxies: starburst -- galaxies: stellar content}

\section{Introduction} 
\label{sec:intro}
 The discovery of an intriguing population of compact, red objects in the early Universe by the James Webb Space Telescope (JWST; \citealt{gardner2023PASP..135f8001G}), termed ``Little Red Dots" (LRDs), has created new opportunities to advance our understanding of galaxy formation and active galactic nuclei (AGN) \citep{furtak2023ApJ...952..142F,
 kokorev2023ApJ...957L...7K,
 kocevski2023ApJ...954L...4K, 
 harikane2023ApJ...959...39H,
 larson2023ApJ...953L..29L,
 labbe2023Natur.616..266L,
 greene2024ApJ...964...39G,
 furtak2024Natur.628...57F,
 matthee2024ApJ...963..129M, 
 barro2024ApJ...963..128B,
 kocevski2025ApJ...986..126K,
 taylor2025ApJ...986..165T,
 labbe2025ApJ...978...92L,
 akins2025ApJ...991...37A}. LRDs exhibit a perplexing combination of spectral properties, including a blue UV continuum, red optical continuum, broad Balmer lines, significant Balmer breaks, limited X-ray detections, and little evidence for time variability. The physical nature of sources that can simultaneously display these observational signatures is still unclear, with ongoing debate over whether they are extreme galaxies \citep{baggen2024ApJ...977L..13B, 
 kokorev2024ApJ...968...38K, 
 perez-gonzalez2024ApJ...968....4P}, a combined signature of compact galaxy and broad-line AGN \citep{guia2024RNAAS...8..207G, 
 kocevski2025ApJ...986..126K, 
 Killi2024A&A...691A..52K, 
 labbe2025ApJ...978...92L, 
 akins2025ApJ...980L..29A, 
 rinaldi2025ApJ...992...71R, 
 hviding2025A&A...702A..57H, 
 leung2025ApJ...992...26L, 
 Marszewski_2026}, supermassive black holes (SMBHs) enshrouded by a gas cocoon \citep{naidu2025arXiv250316596N, 
 deGraaff2025A&A...701A.168D,
 tanaka2025ApJ...995...21T, 
 Schindler2025NatAs.tmp..191S, 
 taylor2025ApJ...989L...7T, 
 rusakov2026Natur.649..574R, 
 mathee2026arXiv260317667M}, and/or evidence of extreme massive black hole accretion \citep{lambrides2025arXiv250909607L, lambrides2026NatAs.tmp...66L}. Regardless of interpretation, their extreme luminosities, compact sizes, and high number densities pose significant challenges for current theoretical models of early galaxy and black hole (BH) growth, not least because their putative BH masses appear greatly overmassive relative to the local relations \citep{durodola2025ApJ...985..169D}. Pinning down how LRDs form and evolve has therefore become central to understanding the interplay between BH and galaxy growth in the first billion years. \\

\indent In AGN scenarios, LRDs are pictured as compact galaxies hosting an active SMBH, with the central engine producing the red optical slope and scattered light feeding the blue UV continuum. Here the broad Balmer lines trace the broad line region (BLR) of BHs with masses $M_{\rm BH}\sim10^{6-8.5}\,\rm M_{\odot}$. This picture is not without tension. The absence of spectral signatures of dust heating in an AGN torus weakens the dust-reddened interpretation \citep{setton2025ApJ...991L..10S, casey2025ApJ...990L..61C}, and the weak or absent variability of LRDs is difficult to reconcile with a uniformly AGN origin (\citealt{furtak2025A&A...698A.227F, zhang2025ApJ...985..119Z, Kokubo2025ApJ...995...24K}; but see \citealt{lambrides2026glimmirspectroscopicvariabilityz7}). Ultra-deep \textit{Chandra} observations likewise return weak or non-detections of X-ray emission, pointing either to a non-AGN origin for some LRDs or to AGN with lower luminosities and/or BH masses than typically inferred \citep{tonima2024ApJ...969L..18A, yue2024ApJ...974L..26Y, Sacchibogdan2025ApJ...989L..30S, lambrides2026NatAs.tmp...66L}. Indeed, several studies argue that LRD bolometric luminosities and inferred SMBH masses may be overestimated by orders of magnitude \citep{lupi2024A&A...689A.128L, greene2026ApJ...996..129G, lambrides2026NatAs.tmp...66L}, in which case a more modest column density would suffice to suppress the X-rays \citep{Sacchibogdan2025ApJ...989L..30S}. \\

\indent A complementary line of work attributes at least part of the LRD population to an exceptionally dense stellar component confined within a compact region \citep{guia2024RNAAS...8..207G, baggen2024ApJ...977L..13B, perez-gonzalez2024ApJ...968....4P, Kokubo2025ApJ...995...24K}. For instance, \citet{baggen2024ApJ...977L..13B} suggested that the Balmer line broadening in some LRDs could arise from galaxy-scale kinematics in massive systems rather than from an AGN BLR. Such purely stellar interpretations are, however, in tension with the ionization lines and related signatures reported by \citet{lambrides2025arXiv250909607L} and \citet{Hviding2026ApJ..1000L..18H}. Spectral energy distribution (SED) modeling more often favors a mixture of stellar light and a central AGN (\citealt{durodola2025ApJ...985..169D, brooks2025ApJ...986..177B, Whalen2025arXiv250921236W, perez-gonzalez2026arXiv260220247P}; but see \citealt{Carranza-Escudero2025ApJ...989L..50C}), and the morphological analysis of \citet{Whalen2025arXiv250921236W} suggests that the LRD population is heterogeneous: a dominant AGN-like subset alongside a smaller ($\sim$15\%) compact-starburst subset. Where broad Balmer lines are absent altogether, as in the sample of \citet{ZhangZijian2026ApJ...998..170Z}, the implication is either the lack of a BLR (and hence of an SMBH) or super-Eddington accretion at the low-mass end of the SMBH distribution. The super-Eddington channel is attractive precisely because it can account for the X-ray weakness, the muted variability, and the apparent overestimation of BH masses within a single framework \citep{madau2024ApJ...976L..24M, 
liu2025ApJ...994..113L, 
inayoshi-maiolino2025ApJ...980L..27I, 
Inayoshi2025ApJ...988L..22I, 
zhang2025ApJ...995...26Z, 
Madau2026arXiv260222386M}. \\

\indent A further set of models does not require a stellar component. In one variant, LRDs are AGN embedded in an extended dusty medium that alone reproduces the unusual SED \citep{Li2025ApJ...980...36L, brooks2025ApJ...986..177B}; in another, they are exotic sources cloaked in a dense, nearly dust-free gas envelope that generates the Balmer break \citep{naidu2025arXiv250316596N, deGraaff2025A&A...701A.168D, taylor2025ApJ...989L...7T, rusakov2026Natur.649..574R, ji2026MNRAS.545f2235J}. Still other proposals invoke direct collapse BHs caught in a luminous phase briefly after formation \citep{Natarajan2024ApJ...960L...1N,cenci2025MNRAS.tmp.1302C, baggen2026ApJ..1002L...4B, pacucci2026arXiv260114368P}, quasi-stars following the core collapse of supermassive stars, with Balmer lines broadened by electron scattering (\citealt{Begelman2026ApJ...996...48B}; but see \citealt{Brazzini2025MNRAS.544L.167B}), low-mass BH seeds captured by a protobulge and accreting at hyper-Eddington rates from hypermagnetized disks \citep{Shi2024ApJ...969L..31S}, early star clusters assembled after a compaction phase \citep{dekel2025arXiv251107578D}, super-Eddington accreting black holes \citep{madau2024ApJ...976L..24M, Pacucci_2024, zhang2025ApJ...995...26Z, lambrides2026NatAs.tmp...66L}, the progenitors of present-day globular clusters \citep{Chisholm2026ApJ..1004L...4C}, or massive galaxies with dust-obscured AGN \citep{Herrero-Carrion2026MNRAS.547ag478H,lachance2026OJAp....955493L}. Despite this proliferation of ideas, the nature of these sources remains elusive \citep{inayoshi-ho2025arXiv251203130I}. \\

\indent Recently, \citet{Marszewski_2026} used high-resolution cosmological zoom-in simulations from the Feedback In Realistic Environments\footnote{\url{http://fire.northwestern.edu}} (\FIRE{}) project \citep{hopkins2018MNRAS.480..800H} to ask whether bursty early galaxies could host the abundant LRD population under AGN scenarios, given that stellar feedback repeatedly evacuates the nuclear gas reservoir and suppresses BH growth \citep{angles-alcazar2017MNRAS.472L.109A, byrne2023MNRAS.520..722B}. Adopting gravitational torque-driven and free-fall accretion with efficiencies that yield BH-to-stellar mass ratios $M_{\rm BH}/M_{\star}\sim0.01$, the \FIRE{} simulations can overproduce the observed abundance of low-luminosity AGN at early times. Notably, restricting the sample to super-Eddington accreting BHs in galaxies with $M_\star > 2\times10^{7}\,\rm M_{\odot}$ simultaneously matches the shape of the observed LRD bolometric and host UV luminosity functions, offering a plausible galaxy-evolution framework for super-Eddington accretion. \\

In this paper, we present complementary work using \FIRE{} simulations to evaluate the viability and limitations of stellar-only interpretations of LRDs. Specifically, we ask which inferred LRD properties can arise in massive, ultra-compact galaxies formed self-consistently from cosmological initial conditions, and which require an additional AGN component. We identify transient compact phases in which simulated galaxies develop small effective radii, high stellar mass surface densities, and extreme dynamical velocities that could help explain the compact sizes, dense stellar components, and broad Balmer lines inferred for LRDs. We then perform dust radiative transfer calculations to generate synthetic SEDs, enabling direct comparison of their continuum properties with observed LRDs, including Balmer break strengths and UV/optical slopes. Finally, we forward model the simulations to produce synthetic emission-line integral field unit (IFU) data cubes for multiple tracers, which we use to assess whether galaxy-scale gas dynamics can reproduce the observed Balmer line widths.

\indent In the following sections, we outline our methodology (Section \ref{sec:methods}), compare the simulated massive galaxies with observed LRDs (Section \ref{sec:results}), and discuss our results and summarize the conclusions of this work (Section \ref{sec:discussion-conclusion}).

\section{Methodology}
\label{sec:methods}
\subsection{FIRE simulations}
\label{subsec:FIRE}
\quad The simulations analyzed here are part of the \FIRE{} project \citep{hopkins2014MNRAS.445..581H, hopkins2018MNRAS.480..800H, hopkins2023MNRAS.519.3154H}. Specifically, we use the \FIREtwo version of galaxy formation physics, with all numerical implementation details fully described in \citet{hopkins2018MNRAS.480..800H}. The simulations use the GIZMO\footnote{\url{http://www.tapir.caltech.edu/~phopkins/Site/GIZMO.html}} code \citep{gizmo2015MNRAS.450...53H}, with hydrodynamics solved using the mesh-free Lagrangian Godunov ``MFM'' method \citep{gaburov2011MNRAS.414..129G, hopkins2018MNRAS.480..800H}. Both hydrodynamic and gravitational (force-softening) spatial resolutions are set in a fully adaptive Lagrangian manner for the gas, where the mass resolution is fixed. The simulations include cooling and heating from a meta-galactic background \citep{FG2009ApJ...703.1416F} and local stellar sources from $T\sim10-10^{10}\,$K; star formation in locally self-gravitating, dense (n$_{\mathrm{H}}\,>$\,1000\,cm$^{-3}$), self-shielding molecular, Jeans-unstable gas; and stellar feedback from OB \&\ AGB mass-loss, SNe Ia \&\ II, and multi-wavelength photo-heating and radiation pressure; with inputs taken directly from stellar evolution models. The \FIRE{} physics, source code, and all numerical parameters are {\em exactly} identical to those in \citet{hopkins2018MNRAS.480..800H}.\\
\indent In this paper, we explore the A, B, and C suites of halos first presented in \citet{massiveFire2016MNRAS.458L..14F,feldmann2017colours} using the \FIREone version and later re-run with the updated \FIREtwo code \citep{ angles-alcazar2017MNRAS.472L.109A,
cochrane2019MNRAS.488.1779C,
cochrane2022ApJ,
cochrane2023MNRAS.523.2409C, 
cochrane2024ApJ...961...37C}. The halos studied here are part of the FIRE-2 public data release\footnote{\url{http://flathub.flatironinstitute.org/fire}} \citep{Wetzel2023,Wetzel2025} and have halo masses as follows (stated for descendants at $z=2$): the A-series of four halos (A1, A2, A4, A8), with $M_{\rm halo}\sim10^{12.5}$ M$_{\odot}$; the B-series of two halos (B1 \& B2), with $M_{\rm halo}\sim10^{13}$ M$_{\odot}$; and the C-series of two halos (C1 \& C2), with $M_{\rm halo}\sim10^{13.5}$ M$_{\odot}$. Halos are identified using the Amiga Halo Finder \citep{knollmann2009ApJS..182..608K} and defined according to the \citet{bryan-norman1998ApJ...495...80B} overdensity definition. Stellar mass growth histories of all halos are presented in Appendix~\ref{sec:allhalos}. The baryonic mass resolutions of the different halos are: $3.3 \times 10^4$\,\msun{} (A-series), $2.7 \times 10^5$\,\msun{} (B-series), and $2.2 \times 10^6$\,\msun{} (C-series), for both gas and star particles. The gravitational softening lengths for the different particles types are: $\epsilon_{\rm gas}$ = 0.7\,pc, $\epsilon_\star$ = 7\,pc, $\epsilon_{\rm BH}$ = 7\,pc, and $\epsilon_{\rm DM}$ = 57\,pc; where $\epsilon_{\rm gas}$ is the minimum adaptive gas softening length and the others are fixed at $z < 9$ in physical units.

\indent We note that these \FIREtwo simulations do not include AGN feedback. The B and C suites do not include BH particles, while the A-suite includes BH growth but no BH feedback \citep{angles-alcazar2017MNRAS.472L.109A}. Black holes in the A-suite do not reach masses high enough to perturb the host galaxy dynamics on the scales analyzed here and therefore their inclusion does not produce meaningful differences relative to the B- and C- suites. Additionally, the absence of AGN feedback in these simulations does not represent a fundamental limitation, since we are exploring the extent to which galaxy-only scenarios can reproduce LRD-like observables, but we discuss the caveats and plausible implications of neglecting AGN feedback in Section \ref{sec:discussion-conclusion}. Throughout this work, we assume a standard flat $\Lambda$ cold dark matter cosmology with $h$ $\approx$ 0.7, $\Omega_{\mathrm{M}}=1-\Omega_{\mathrm{\Lambda}}\approx0.27$, and $\Omega_{\mathrm{b}}\approx$\,0.045 \citep{planck2014A&A...571A..16P}. We adopt a \cite{kroupa2001MNRAS.322..231K} stellar initial mass function.

\subsection{Synthetic continuum observations}
\label{subsec:SED-methods}
The synthetic SEDs are generated using the methods described in \cite{cochrane2019MNRAS.488.1779C, cochrane2023MNRAS.523.2409C, cochrane2024ApJ...961...37C}. In summary, the Monte Carlo dust radiative transfer code \SKIRT \citep{baes2011ApJS..196...22B} is used to make predictions for the ultraviolet to far-infrared SED. Dust is modeled as tracing metal-rich gas, with a dust-to-metals mass ratio of $0.4$ and a \cite{Weingartner2001} Milky Way prescription for its composition. Gas above 10$^6$\,K is assumed to be dust-free to account for thermal sputtering of grains. The radiation transfer is performed on an octree grid, the cell sizes of which are adjusted so that no more than 0.0001 per cent of the total dust mass is contained in one cell. In order to probe a range of sightlines, we generate synthetic SEDs for each simulated galaxy along $7$ lines of sight (rotating the galaxy $0,30,60,90,120,150$ and $180$ degrees with respect to the ``face-on'' orientation, which is determined by the angular momentum vector of the gas particles). Further details of the dust modeling and dust attenuation curve properties of these simulated galaxies are provided in \cite{cochrane2024ApJ...961...37C}. \\
\indent In this paper, we explore SEDs for snapshots where simulated galaxies have sizes consistent with observed LRDs. For each selected snapshot, we compute the SED including all stars and dust within $<3$\,kpc, which is larger than the stellar half-mass radius $R_{\mathrm{eff}}$ of the LRD candidates that we explore. We find that photons from the stellar population within $R_{\mathrm{eff}}$ dominate the SEDs in the rest frame UV to Near-IR wavelength range, so this is a sufficiently large aperture. We compute Balmer break strengths from the SEDs of \FIRE{} galaxies using a method following previous work \citep{Binggeli2019MNRAS.489.3827B, Roberts-Borsani2024ApJ...976..193R}, where the strength is defined as the ratio of fluxes $F_{\upnu}$ at 4200\,\r{A} (red flux) and 3500\,\r{A} (blue flux): 

\begin{equation}
    \rm Balmer\ break\ strength = \frac{\mathit{F}_\upnu(4200\,Å)}{\mathit{F}_\upnu(3500\,Å)}.
    \label{eq:BBS}
\end{equation}

For the selected ultra-compact galaxy phases, we also calculate the absolute UV magnitude ($M_{\rm UV}$) in the AB system, leveraging synthetic SEDs to compare with observed UV luminosity functions. We first convert the flux density at wavelength 1500\,\r{A} into apparent AB magnitude as:

\begin{equation}
    m_{\rm AB} = -2.5 \,\rm log_{10}\left(\frac{\mathit{f}_{\rm UV}}{3631\,Jy}\right),
    \label{eq:apperent mag}
\end{equation}

\noindent which is normalized for the zero point magnitude \citep{oke1983ApJ...266..713O, tonry2012ApJ...750...99T}. Then we convert $m_{\rm AB }$ to absolute magnitude $M_{\rm UV}$ using:

\begin{equation}
    M_{\rm UV} = m_{\rm AB} - 5\,\rm log_{10}\left(\frac{\mathit{d}_L}{10\,pc}\right),
    \label{eq:absolute mag}
\end{equation}

\noindent where $d_{\rm L}$ is the luminosity distance in pc for the corresponding galaxy redshift.

\begin{figure*}
    \centering
    \includegraphics[width=0.98\linewidth]{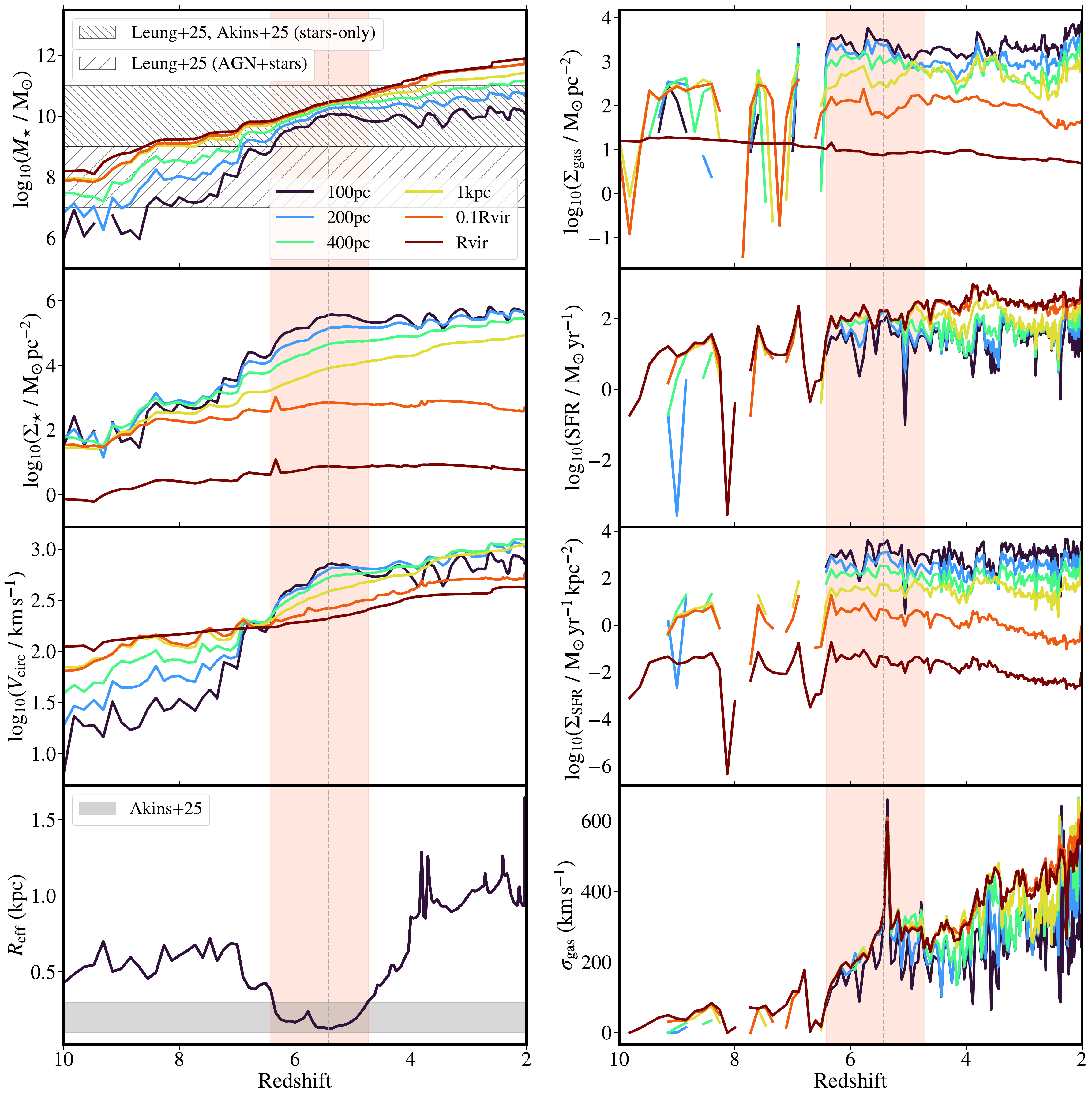}
    \caption{Redshift evolution of different physical properties of galaxy B2 calculated within 100\,pc (black), 200\,pc (blue), 400\,pc (green), 1\,kpc (yellow), 0.1\,$R_{\mathrm{vir}}$ (orange), and $R_{\mathrm{vir}}$ (brown) from the center. The horizontal hatched bands in the $M_{\star}$ panel indicate the approximate range of LRD stellar masses inferred in the stellar-only \citep{leung2025ApJ...992...26L, akins2025ApJ...991...37A} and AGN+stellar \citep{leung2025ApJ...992...26L} scenarios, roughly separating at $M_\star\sim10^9$\,M$_{\odot}$. The horizontal band in the $R_{\rm eff}$ panel shows the estimated effective radii range of LRDs from \citet{akins2025ApJ...991...37A}. The vertical band represents the range of redshifts when the simulated galaxy is as compact as observed LRDs, with $R_{\mathrm{eff}}\,<\,300\,\mathrm{pc}$. The vertical dashed gray line represents the redshift at which $R_{\mathrm{eff}}$ is the lowest; we further investigate properties of the galaxy during this epoch in Figures \ref{fig:physical-prop}--\ref{fig:RGBs}. After an initial early phase of bursty star formation with stellar feedback repeatedly evacuating gas from the galaxy ($z \gtrsim 6.5$), a strong central starburst with $\mathit\Sigma_{\mathrm{SFR}}>$\,1000 M$_{\odot}\,$yr$^{-1}\,$kpc$^{-2}$ drives the formation of an ultra-dense stellar system reaching stellar mass surface density $\mathit\Sigma_{\mathrm{\star}}$ $\sim 10^{5.5}$\,M$_{\odot}\,\mathrm{pc}^{-2}$ and circular velocity $V_{\mathrm{circ}}\sim800$\,\kmps{} in the central 100\,pc. During this ultra-compact phase, which lasts $\sim$400\,Myr, galaxy B2 has $M_{\star}\sim10^{10}\,$\msun{} and $R_{\rm eff} < 300\,\rm pc$, consistent with observational constraints of LRDs in stellar-only scenarios.
    }
    \label{fig:redshift-evolution}
\end{figure*}

\subsection{Synthetic emission line observations}

Synthetic emission line data are generated for the simulated galaxies following the methodology described in \cite{Roy2026}, using a pipeline combining the chemistry solver {\sc{chimes}} \citep{richings14a2014MNRAS.440.3349R, richings2014MNRAS.442.2780R} and the 3D Monte Carlo radiative transfer code {\sc{radmc-3d}} \citep{dullemond2012}. {\sc{chimes}} calculates the ion abundances of 157 different species of atoms, ions, and molecules, considering photo-ionization from young stars as well as local shielding of ionizing photons by gas. {\sc{chimes}} ion abundances are projected onto an adaptive mesh refinement (AMR) grid along with other physical properties such as gas density, gas velocity, stellar mass, and dust mass, which are used to perform the line radiative transfer with {\sc{radmc-3d}}. We create synthetic IFU data cubes for [\ion{C}{2}]\,158\,\microm{}, H$\upalpha$, H$\upbeta$, [\ion{N}{2}]\,6585\,\r{A}, and [\ion{O}{3}]\,5007\,\r{A} emission lines. Except for Balmer lines, level populations are calculated based on collisional excitation only. For the Balmer lines, recombination is also modeled based on a cascade matrix formalism \citep{Raga2015, rihchings2021MNRAS.503.1568R}. In recent work, \citet{Liu2026MNRAS.547ag429L} has shown that the method \cite{Raga2015} use may underestimate the Balmer line luminosities by a factor of $\sim$2, but this does not affect the conclusions of our work, as we are primarily interested in the line widths and kinematics of the Balmer lines, which are unaffected, rather than the overall luminosity.

We model the impact of dust absorption and scattering on the
emission line observables by assuming two species of dust grains -- silicate and graphite. During the emission line modeling, we take into account the depletion of metals on dust grains, which impacts the gas-phase metallicity and in turn affects the estimated amount of dust itself.\footnote{Although this dust prescription differs from that used in SKIRT, its impact is negligible here because the synthetic IFUs cover very narrow wavelength ranges and are used only to analyze emission-line gas kinematics.} The depletion factors of individual elements are adopted from \citet{decia2016A&A...596A..97D}, which extended the dust depletion model of \citet{jenkins2009ApJ...700.1299J}. As in SKIRT, we implement a temperature threshold ($T>10^6$\,K) in our emission line modeling to identify dust-free gas due to sputtering of dust grains. 
More details of the dust model used in this study can be found in \citet{richings2022MNRAS.517.1557R}.  

\indent The intrinsic spectral resolution of the IFUs is set to 5\,\kmps{} so that they can capture the local turbulence-driven broadening from gas in individual AMR cells, set to a physically motivated value of 7.1\,\kmps \citep{rihchings2021MNRAS.503.1568R}. To roughly mimic the spectral resolution of the mid-resolution grating of JWST NIRSpec, we first convolve each IFU image slice with a Gaussian kernel corresponding to a spatial resolution of 0.4 kpc, and subsequently convolve the spatially integrated spectra with a Gaussian kernel of $\rm FWHM=300\,$\kmps{}. However, we note that among the emission lines, [\ion{C}{2}]\,158\,\microm{} is outside the wavelength range of JWST even though is convolved here in a similar way as the other rest frame optical lines.

\section{Results}
\label{sec:results}

In the following, we compare the physical properties of the simulated \FIREtwo galaxies to those inferred from observations of LRDs.

\begin{figure*}
    \centering
    \includegraphics[width=1\linewidth]{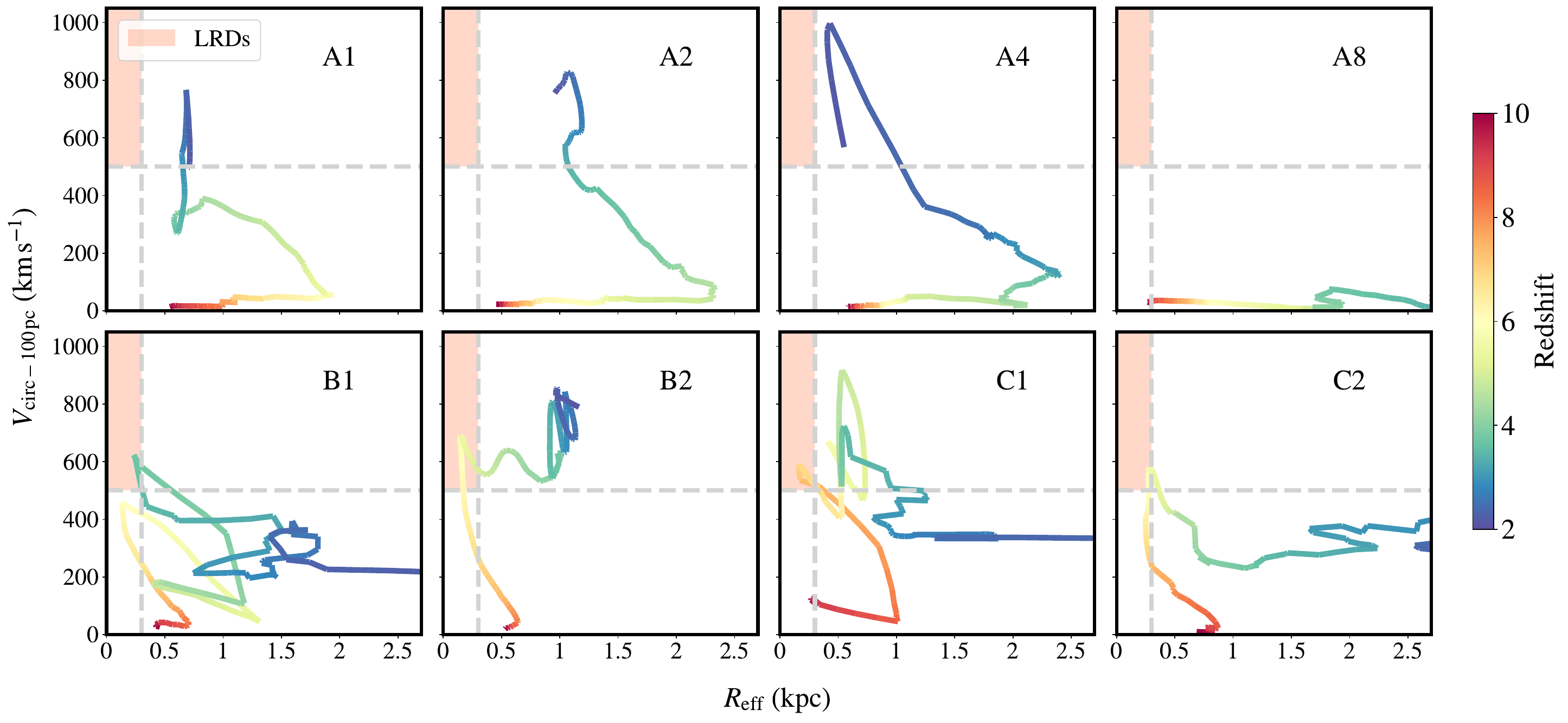}
    \caption{Evolutionary trajectories of simulated galaxies in the circular velocity ($V_{\mathrm{circ}}$; measured within 100\,pc)--stellar half mass radius ($R_{\mathrm{eff}}$) plane, color coded by redshift. The shaded region indicates the parameter space occupied by observed LRDs: $R_{\rm eff}<300\,\rm pc$ and broad Balmer line width $>1000\,$\kmps{} (interpreted as galaxy-scale dynamics with $V_{\rm circ} > 500$\,\kmps). 
    Galaxies in dark matter halos selected to reach $M_{\mathrm{halo}}\,\gtrsim\,10^{13}$\,M$_{\odot}$ at $z = 2$ (halos B1, B2, C1, and C2) exhibit ultra-compact, LRD-like phases in the redshift range $z \approx 4-8$, while halos below this mass (A1, A2, A4, and A8; selected to reach $M_{\mathrm{halo}}\approx10^{12.5}$\,M$_{\odot}$ at $z=2$) do not become compact enough at early times. In our model, ultra-compact galaxies that feature LRD-like properties are central galaxies residing in the progenitors of group-size halos at $z=0$ ($M_{\mathrm{halo}}$ $\gtrsim$ 10$^{13.5}$\,M$_{\odot}$).
    }
    \label{fig:vcirc-vs-reff-hist2d}
\end{figure*}

\begin{figure*}
    \centering
    \includegraphics[width=\linewidth, trim = {1.2cm 0.5cm 0.5cm 0.38cm}]{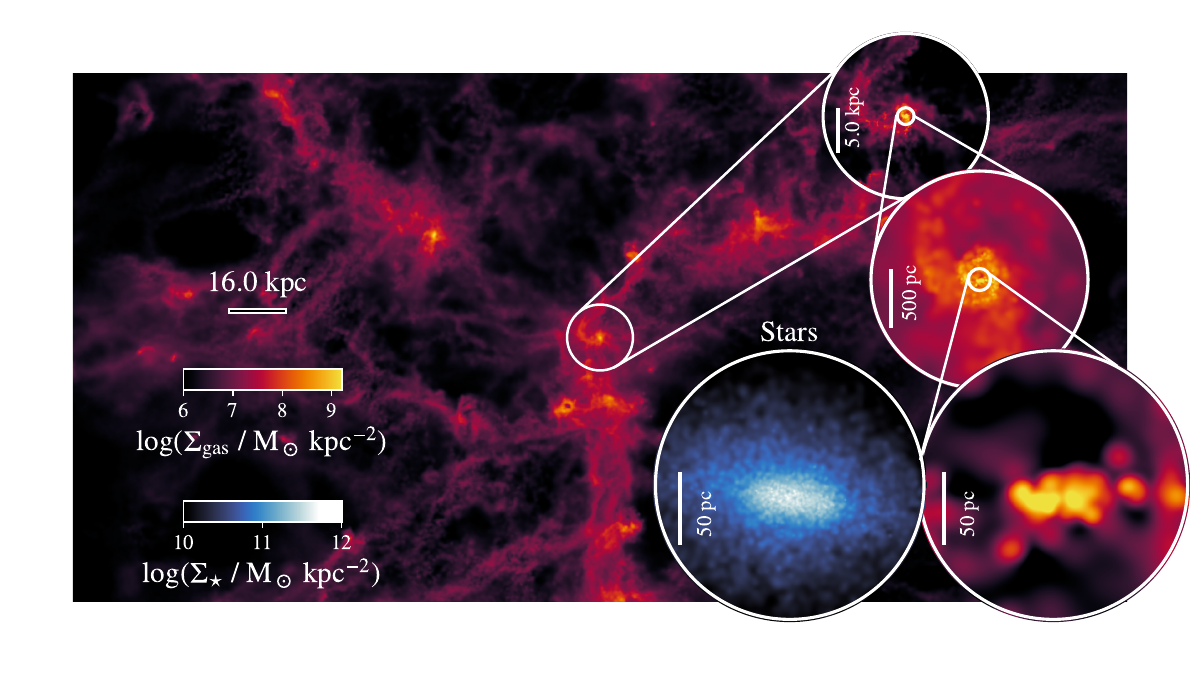}
    \caption{Visual illustration of the gas mass surface density across spatial scales for galaxy B2 during a representative ultra-compact phase at $z\sim5.4$. The stellar mass surface density is also shown for the central $100\,\rm{pc}$ region, which encloses $M_{\star}>10^9\,$M$_{\odot}$. The background image shows the complex, large-scale environment of the simulated LRD candidate, which is surrounded by several galaxies embedded in a cosmic web filament. Tidal tails and gas streamers connect interacting galaxies, while the high gas inflow rate ($\dot{M}\gtrsim10^3\,\rm M_{\odot}yr^{-1}$) drives the formation of a turbulent, ultra-dense nuclear gas disk and the subsequent development of an ultra-compact stellar system in the inner $100\,\rm{pc}$. }
    \label{fig:physical-prop}
\end{figure*}

\begin{figure*} 
    \centering 
    \includegraphics[width=0.49\linewidth,
        trim={0.9cm, 0.5cm, .7cm, 0.5cm}]{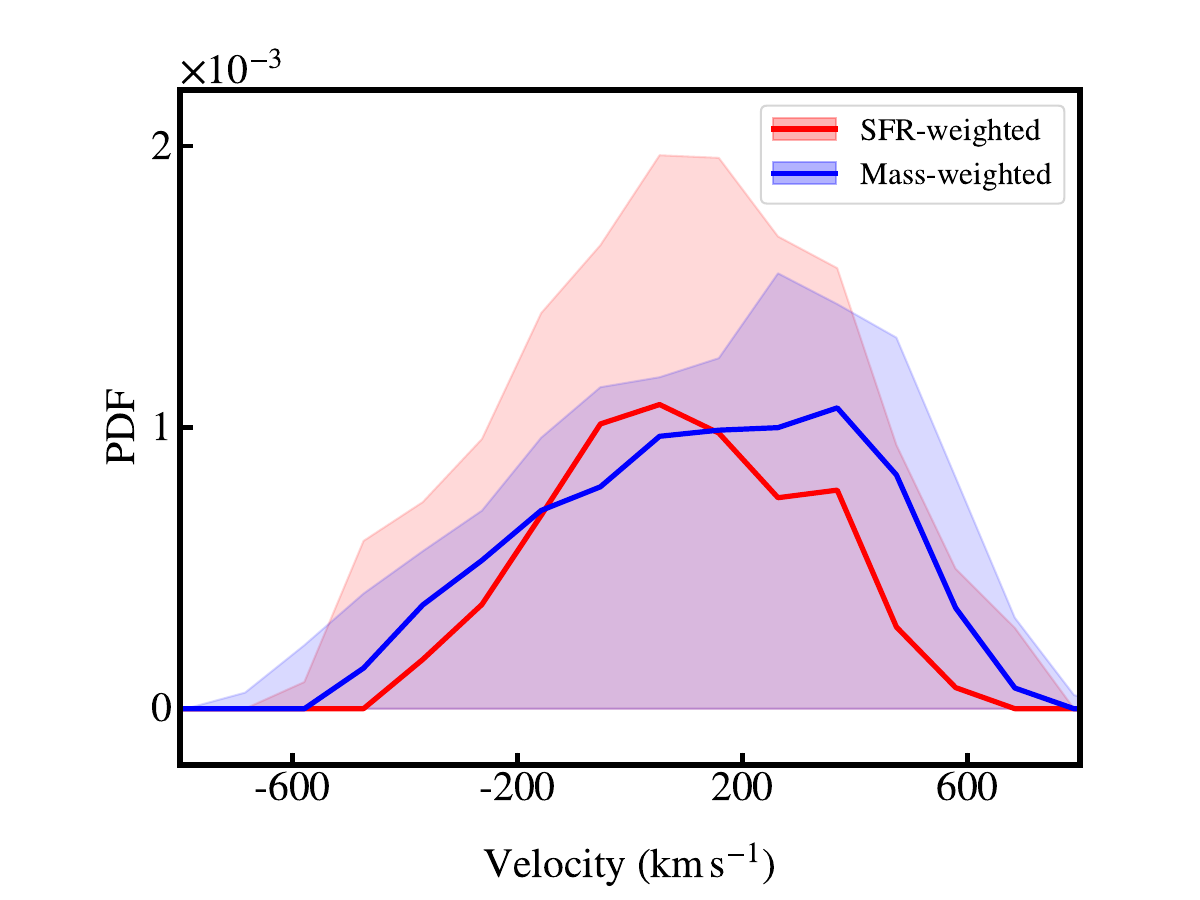}
    \hfill
    \includegraphics[width=0.49\linewidth,
        trim={0.9cm, 0.5cm, .7cm, 0.5cm}]{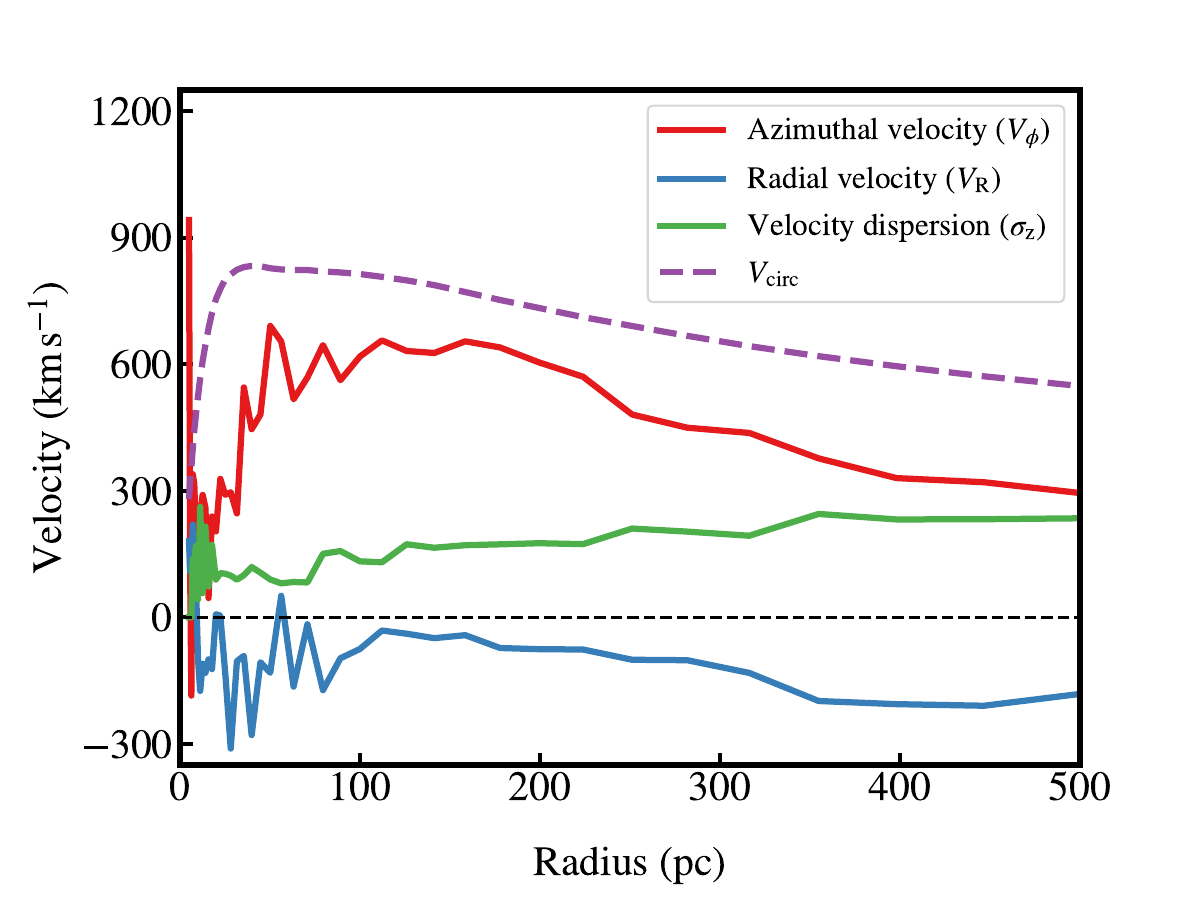}
    \caption{Kinematic properties of gas in galaxy B2 during a representative ultra-compact phase at $z\sim5.4$. Left: Probability density functions of SFR-weighted (red) and mass-weighted (blue) line of sight (LOS) gas velocity, where we consider 100 random sight-lines through the central $100$\,pc of the galaxy. The solid lines represent median values and the shaded regions indicate 16$^{\rm th}$--84$^{\rm th}$ percentile ranges. The kinematic similarity between the mass-weighted and SFR-weighted LOS velocity profiles suggests that star formation is occurring across the full dynamic range of the gas reservoir. Right panel: Azimuthal velocity (red), radial velocity (blue), velocity dispersion along the z-axis (green), and total circular velocity (purple dashed) as a function of radial distance. With the azimuthal velocity $V_{\upphi} > 600$\,\kmps{} reflecting the extreme gravitational potential, the bulk motion of gas on $\sim$100\,pc scales could drive the broadening of emission lines to $\rm FWHM\gtrsim 1200$\,\kmps, potentially mimicking the inferred BLR kinematics of observed LRDs in Balmer lines.}
    \label{fig:sfr-weighted-hist-and-rotation}
\end{figure*}

\subsection{Redshift evolution of physical properties}

Following this ultra-compact phase, the galaxy gradually expands. The effective radius increases while central surface densities stay similar, implying that subsequent star formation and dynamical processes accumulate stellar mass outward. This structural evolution suggests that this ultra-compact, LRD-like phase represents a transient but natural stage in the lifecycle of massive galaxies at high redshift.

We begin by exploring the redshift evolution of different physical properties of the simulated galaxy B2 to identify potential evolutionary phases exhibiting some of the key properties of LRDs.
Figure~\ref{fig:redshift-evolution} presents the following properties computed for different radial apertures:
stellar mass ($M_\star$), stellar mass surface density ($\mathit \Sigma_\star$), circular velocity ($V_{\rm circ}$), stellar half mass radius ($R_{\rm eff}$),
gas mass surface density ($\mathit \Sigma_{\rm gas}$), total instantaneous star formation rate (SFR), SFR surface density ($\mathit \Sigma_{\rm SFR}$), and SFR-weighted gas velocity dispersion ($\sigma_{\rm gas}$). Here, $R_{\rm eff}$ is defined as the half-mass radius of all stars within 10\% of the virial radius and $\mathit \sigma_{\rm gas} = \sqrt{(\sigma_x^2+\sigma_y^2+\sigma_z^2)/3}$, where $\sigma_x,\, \sigma_y,$ and $\sigma_z$ are the one-dimensional SFR-weighted gas velocity dispersion along the x, y, and z axes. 
A subset of these properties are presented for all simulated halos in Appendix~\ref{sec:allhalos}, but here we focus on B2 as a representative example. 

At early times ($z \gtrsim 6.5$), galaxy B2 undergoes bursty star formation regulated by strong stellar feedback. This phase is characterized by large fluctuations in the central gas surface density, SFR, and velocity dispersion, indicating repeated episodes of gas inflow and feedback-driven evacuation. Stellar mass growth during this stage is modest within the innermost 100–400 pc, and the system remains relatively extended. A qualitative transition occurs near $z\sim6$, when a sustained central starburst is triggered. During this episode, the gas surface density in the central 100\,pc rises sharply, leading to extreme star formation rate surface densities ($\mathit\Sigma_{\rm SFR} > 10^3\,{\rm M}_\odot\,{\rm yr}^{-1}\,{\rm kpc}^{-2}$). This intense, centrally concentrated star formation rapidly builds an ultra-dense stellar core, driving the stellar surface density to $\mathit\Sigma_\star \sim 10^{5.5}\,{\rm M}_\odot\,{\rm pc}^{-2}$. Concurrently, the circular velocity within 100\,pc increases dramatically, reaching $V_{\rm circ}\sim800\,{\rm km\,s^{-1}}$, signaling the formation of a deep central potential well dominated by baryonic matter.

During this period, the effective radius contracts to $R_{\rm eff}\sim200$–300\,pc, and this stage persists for $\sim$400\,Myr. The minimum of $R_{\rm eff}$ (indicated by the vertical dashed line) identifies the epoch when the galaxy is maximally centrally concentrated. During this ultra-compact phase, the galaxy simultaneously attains sizes consistent with observed LRDs and stellar masses of $M_\star \sim 10^{9-10}\,{\rm M}_\odot$ within $R_{\rm eff}$, which is in line with stellar-only models \citep{akins2025ApJ...991...37A, leung2025ApJ...992...26L, Carranza-Escudero2025ApJ...989L..50C}. The horizontal bands in the $M_\star$ and $R_{\rm eff}$ panels indicate observational constraints, showing that the simulated system naturally enters the compactness part of the LRD parameter space without fine-tuning.\footnote{We note that the observed compactness of LRDs, typically expressed as an observed half-light radius $R_{\rm eff}\lesssim300$\,pc \citep{akins2025ApJ...991...37A}, does not in general directly constrain the total stellar half-mass radius of the host galaxy. In AGN-dominated interpretations, where much of the rest-optical light originates from a central point source, the galaxy may contain a more extended stellar component that remains below the surface-brightness detection threshold of current imaging. The condition $R_{\rm eff}<300$\,pc identified here for the stellar half-mass radii of our simulated galaxies should therefore be read as sufficient but not necessary for a system to appear as a sub-300\,pc source in JWST imaging. Sources with larger total stellar half-mass radii could satisfy the observed-size criterion if their light is dominated by a compact AGN or by a compact high-surface-brightness stellar component. However, if stars are invoked to explain the UV part of the LRD SED, then the UV-emitting stellar population must itself be confined to the observed rest-UV size scale. The compact stellar distributions in our simulations are thus directly relevant to stellar explanations of the blue UV continuum, while not excluding additional, lower-surface-brightness stellar mass at larger radii.}
Importantly, the compact phase is not merely a consequence of global mass growth. While stellar mass increases smoothly across all apertures, the most dramatic evolution occurs in the innermost 100–400 pc, indicating that dissipative gas inflow and centrally concentrated star formation drive the structural transformation, rather than halo-scale assembly alone. The steep rise in $\mathit\Sigma_\star$ and $V_{\rm circ}$, together with elevated gas velocity dispersion, suggests a rapidly assembled, dynamic nuclear system.

In Figure~\ref{fig:vcirc-vs-reff-hist2d}, we present the evolutionary trajectories of simulated galaxies in the $V_{\text{circ}}$--$R_{\text{eff}}$ plane, which reveal a contrast in structural development driven by host halo mass. Each galaxy's path is color-coded by redshift, and the shaded region indicates the parameter space occupied by observed LRDs, which we use to identify LRD-like phases in simulated galaxies. Here we consider the upper limit to LRD sizes $R_{\rm eff}<300\,\rm pc$ \citep{akins2025ApJ...991...37A} and a minimum circular velocity $V_{\rm circ} > 500$\,\kmps~representative of broad Balmer line widths $>1000\,$\kmps{} under the stellar-only interpretation of galaxy-scale dynamics \citep{baggen2024ApJ...977L..13B}.
Galaxies residing in the most massive progenitors ($M_{\text{halo}} \gtrsim 10^{13} M_{\odot}$ at $z=2$), represented by the B and C series, undergo a rapid ultra-compact phase between redshifts $z \approx 8$ and $z \approx 4$, which interestingly coincides with the redshift range of most observed LRDs \citep[e.g.,][]{kocevski2025ApJ...986..126K}. During this epoch, intense gas inflows funnel significant baryonic mass into the galactic centers, causing a sharp decrease in $R_{\text{eff}}$ while simultaneously driving the central circular velocity $V_{\rm circ} \gtrsim$ 500\,\kmps. This coupled evolution effectively pushes these systems into the LRD-like parameter space (shaded region), characterized by sub-kiloparsec scales and high rotational velocity. 
In contrast, galaxies inhabiting comparatively lower-mass halos ($M_{\text{halo}} \sim 10^{12.5}\,\rm M_{\odot}$ at $z=2$; A-series) fail to achieve the threshold densities required to enter the LRD regime. 
While these systems exhibit a steady increase in $V_\mathrm{circ}$ as they accumulate mass, their $R_\mathrm{eff}$ remains consistently above the $\sim$300\,pc limit during the high-redshift window, a reflection of their lower halo masses and later assembly histories rather than the absence of AGN feedback, which primarily affects the structural evolution of these systems at $z \lesssim 2$ \citep{parsotan2021MNRAS.501.1591P, cochrane2023MNRAS.523.2409C}.
This suggests that ultra-compact galaxies could be transient, high-redshift phases unique to the central galaxies of group-size dark matter halos with $M_{\mathrm{halo}}$ $\gtrsim$ 10$^{13.5}$\,M$_{\odot}$ at $z=0$.

Following the peak compaction phase seen in the more massive hosts (B1, B2, C1, C2), the trajectories generally trend toward the lower-right quadrant of the plane at $z < 4$. This structural relaxation marks a transition to inside-out growth, in which subsequent star formation and dry mergers primarily add mass to the galactic outskirts. Consequently, the ultra-compact LRD-like morphology appears to be a progenitor state that is eventually extended as the galaxy evolves into a more massive system. Among the galaxies we explore in this paper, B2 exhibits the longest LRD-like phase throughout its evolution, during the redshift range $z \sim 4.5$--6.5. We focus some of our analysis below on galaxy B2 given that it offers the largest number of LRD-like snapshots with $V_{\mathrm{circ}}\,>\,500$\,\kmps~and $R_{\mathrm{eff}} < 300\,$pc, but we emphasize that galaxies B2, C1, and C2 undergo qualitatively similar evolution (unlike the A-series halos), as shown in Appendix~\ref{sec:allhalos} (see Figure~\ref{fig:all-massive-halos}).

\subsection{Morphology and kinematics}

A multiscale view of the gas and stellar structure of galaxy B2 can be seen in Figure~\ref{fig:physical-prop} during a representative ultra-compact phase at $z \sim5.4$. On large scales, the system is embedded in a dense, filamentary environment, with diffuse gas streams and nearby companions linked by tidal features and inflowing structures. The gas surface density maps show that the galaxy does not evolve in isolation; instead, the central object is continuously fed by the surrounding cosmic web and by interactions within its local environment. This large-scale inflow appears to drive substantial gas accumulation toward the nucleus (at a rate $\dot{M}\gtrsim10^3\,\rm M_{\odot}yr^{-1}$ within the inner 100\,pc), producing a highly concentrated central reservoir of gas within an otherwise complex and dynamically disturbed halo.

The zoomed panels reveal that this inflow results in the formation of an extremely dense nuclear gas structure on sub-kiloparsec and then sub-100\,pc scales. The gas becomes progressively more compact and clumpy toward the center, while the stellar mass map showing the innermost $\sim100$\,pc already hosts a remarkably dense stellar component reaching $\mathit{\Sigma}_\star > 10^{11}\,\rm M_\odot\,kpc^{-2}$. Together, these panels illustrate a physically coherent picture in which large-scale accretion, tidal interaction, and central gas compression combine to build an ultra-compact stellar system embedded within a dense gaseous core. In the context of LRD formation, Figure~\ref{fig:physical-prop} suggests that the extreme compactness inferred from observations can arise naturally from a hierarchical, inflow-driven evolutionary phase in massive high-redshift galaxies.

Figure~\ref{fig:sfr-weighted-hist-and-rotation} presents a kinematic analysis of galaxy B2 at $z \sim 5.4$,
demonstrating that the ultra-compact stellar core is embedded within an extreme gravitational potential capable of driving broad-line emission through bulk gas motions alone. The left panel shows one-dimensional velocity probability density functions (PDFs) for 100 random lines-of-sight (LOS), revealing a significant velocity spread, with both SFR-weighted and mass-weighted velocity distributions extending to $\pm 600$\,\kmps. This suggests that the gas reservoir is highly dynamic and subject to large-scale coherent rotation, and the similarity between the mass-weighted and SFR-weighted LOS velocity PDFs reveals that the star-forming gas is kinematically well-coupled to the bulk gaseous reservoir. This kinematic similarity suggests that star formation is occurring across the full dynamic range of the gas reservoir, rather than being confined to a low-velocity, settled disk. The mass-weighted distribution exhibits a relatively smooth, broad profile, reflecting the continuous distribution of the interstellar medium (ISM) across the gravitational potential. On the other hand, the SFR-weighted PDF displays an additional off-centered peak at $V_{\rm LOS}\approx400\,$\kmps, likely corresponding to localized, high-density star-forming clumps moving at orbital velocities close to this value. Despite these localized variations, the $16^{\text{th}}$–$84^{\text{th}}$ percentile ranges for both distributions are nearly identical, with both profiles extending to $\pm 600$\,\kmps. 

This picture is further supported by the right panel of Figure~\ref{fig:sfr-weighted-hist-and-rotation}, where we compare different radial velocity profiles to the circular velocity of galaxy B2. The azimuthal/rotational velocity ($V_{\upphi}$) dominates the kinematics, peaking at $V_{\upphi} > 600$\,\kmps{} within the inner $100\,$pc and closely tracing the total circular velocity, which reaches $V_{\text{circ}} \sim 800$\,\kmps~at $\sim$50\,pc. While primarily rotationally supported, the ISM is highly turbulent, with vertical velocity dispersion reaching $\sigma_{\rm z} \sim 200$\,\kmps, and shows net inflow with radial velocity $V_{\rm R} \sim -[100$--200]\,\kmps~across the galaxy.
The presence of such high-velocity, organized rotation of star forming gas, coupled with substantial vertical velocity dispersion, indicates that broadening of Balmer lines up to $\text{FWHM} \approx 2 \times V_{\upphi} \times {\rm sin}(i) \sim 1200$\,\kmps{} (for edge-on view with $i = 90^{\circ}$) could be accounted for by the host's gravitational potential alone rather than requiring an AGN broad-line region.

\begin{figure*}
    \centering 
    \begin{minipage}[t]{0.49\linewidth}
        \centering
        \includegraphics[width=\textwidth,
            trim = {1cm, 1cm, 1cm, 0cm}]{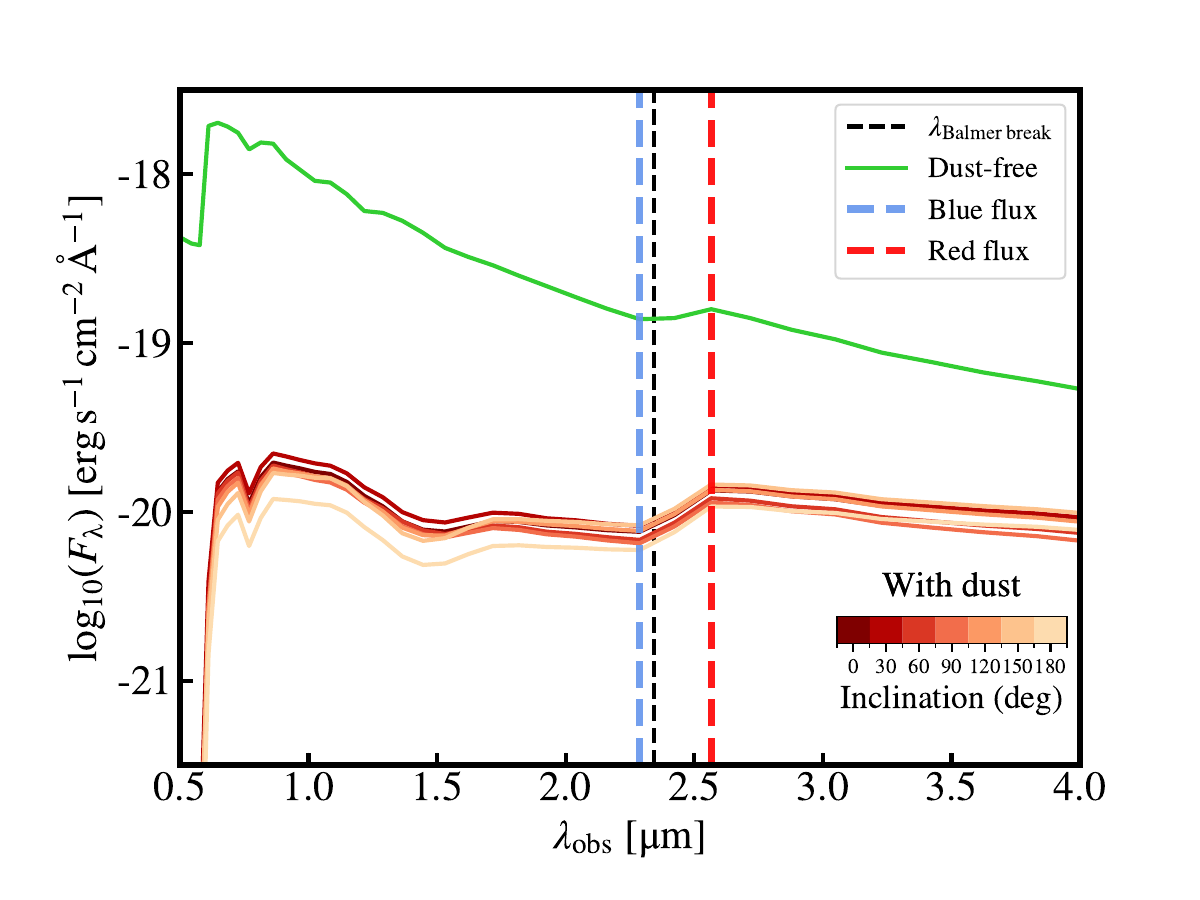}
    \end{minipage}
    \hfill 
    \begin{minipage}[t]{0.49\linewidth}
        \centering
        \includegraphics[width=\textwidth,
            trim = {1cm, 1cm, 1cm, 0cm}]{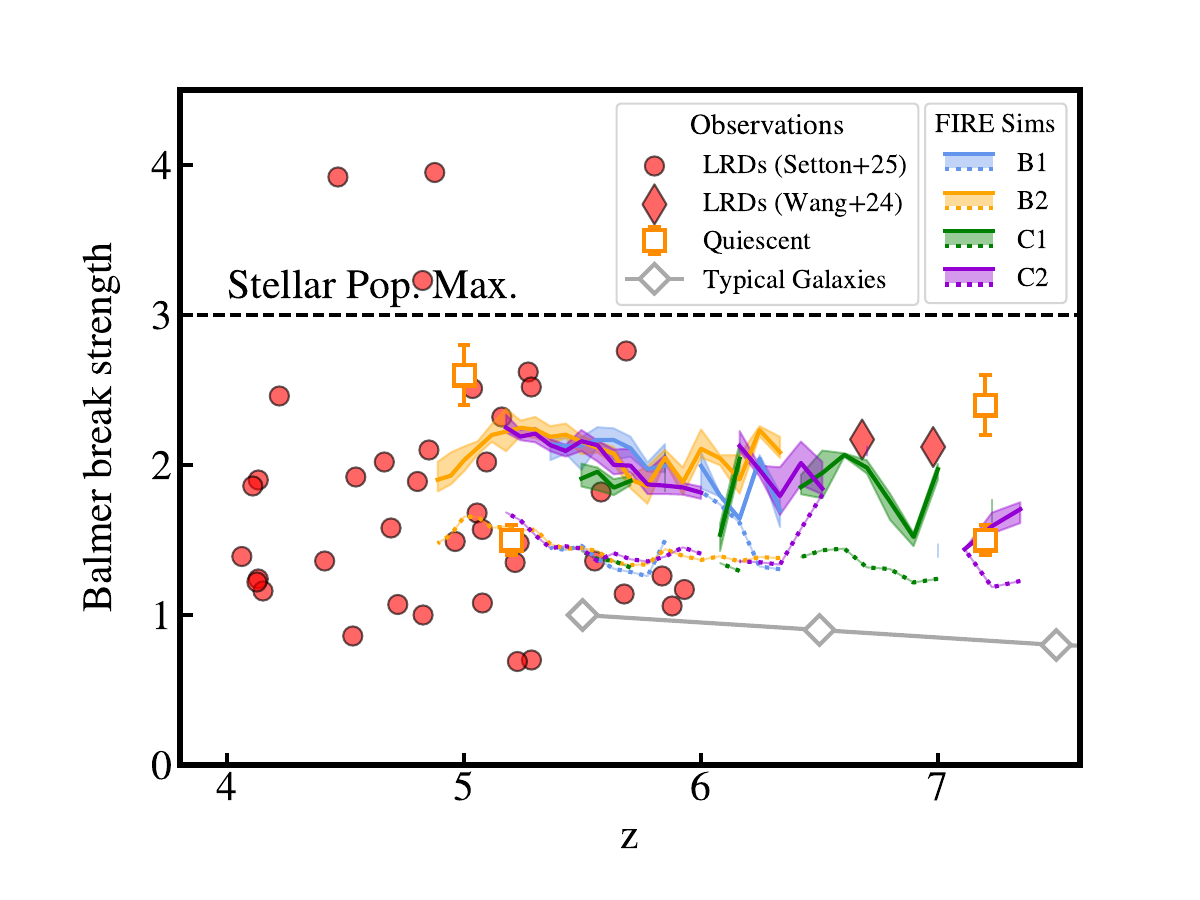}
    \end{minipage}
    \caption{Synthetic SEDs (left) and Balmer break strengths (right) of simulated galaxies for LRD-like phases during which they are both massive and compact ($V_{\mathrm{circ}}\,>\,500$\,\kmps; $R_{\mathrm{eff}} < 300\,$pc). The SED of galaxy B2 at $z\sim5.4$ is shown on the left for different inclination angles (brown-orange color scale). Fluxes ($F_{\upnu}$) at wavelengths indicated by the blue and red vertical dashed lines are used to calculate the Balmer break strength. The SED assuming no dust effects (emission, absorption, or scattering) is shown in green. For all simulated galaxies, the right panel shows the median (solid line) and full range (shaded region) of Balmer break strengths across seven lines-of-sight as a function of redshift, with dotted lines indicating dust-free Balmer break strengths for each galaxy. Simulated ultra-compact galaxies at early times show Balmer break strengths (and negative beta slopes) similar to observed LRDs \citep{setton2025ApJ...995..118S, Wang2025ApJ...984..121W} and high redshift quiescent galaxies \citep{deGraaff2025NatAs...9..280D, weibel2025ApJ...983...11W}, and almost twice that of typical star forming galaxies at similar redshifts \citep{Roberts-Borsani2024ApJ...976..193R}.}
    \label{fig:SEDs}
\end{figure*}

\subsection{Synthetic SEDs and emission line profiles}

Here, we analyze the observable properties of our simulated ultra-compact galaxy phases with $R_{\mathrm{eff}} < 300\,$pc and $V_{\mathrm{circ}}\,>\,500$\,\kmps~to investigate if they naturally reproduce LRD observables during this compact phase. 
The left panel of Figure~\ref{fig:SEDs} presents the SED of galaxy B2 at $z\sim5.4$ in the observed frame for different inclination angles, as shown by the brown color scale. Vertical dashed lines indicate the Balmer break wavelength (black), and the bracketing rest-frame wavelengths 3500\,\r{A} (blue) and 4200\,\r{A} (red) used to compute the Balmer break strength (Equation~\ref{eq:BBS}).
Common spectral features of observed LRDs include a V-shape in their $F_{\lambda}$ profile, with negative rest-frame UV-slope ($\beta_{\rm UV}$) and positive optical slope ($\beta_{\rm opt}$), and prominent Balmer break strengths. Across all inclination angles, the simulated SEDs for galaxy B2 show Balmer break strengths $\sim$2 and negative UV-slopes $\beta_{\rm UV}=-1.24\pm 0.21$, consistent with observed LRDs \citep{setton2025ApJ...995..118S, kocevski2025ApJ...986..126K}. However, the absence of a positive optical slope suggests that our synthetic SEDs require an additional spectral component to fully reproduce an LRD continuum spectrum. 

To assess how much of this spectral shape is driven by dust attenuation versus intrinsic stellar emission, we produce dust-free SEDs explicitly ignoring dust emission, absorption, and scattering (shown as the green solid line; independent of inclination). The dust-free SED shows a weaker Balmer break ($\sim$1.5), steeper negative UV slope, and fluxes $\sim$2 orders of magnitude higher than the fiducial SEDs, confirming that while dust attenuation significantly reshapes the UV--optical continuum and deepens the apparent Balmer break, it does not generate the red optical slope characteristic of LRDs, reinforcing the need for a non-stellar component at rest-frame optical wavelengths.

\begin{figure*}
    \centering
    \includegraphics[width=1\linewidth,
    trim=5cm 3cm 6cm 7cm,
    clip
    ]{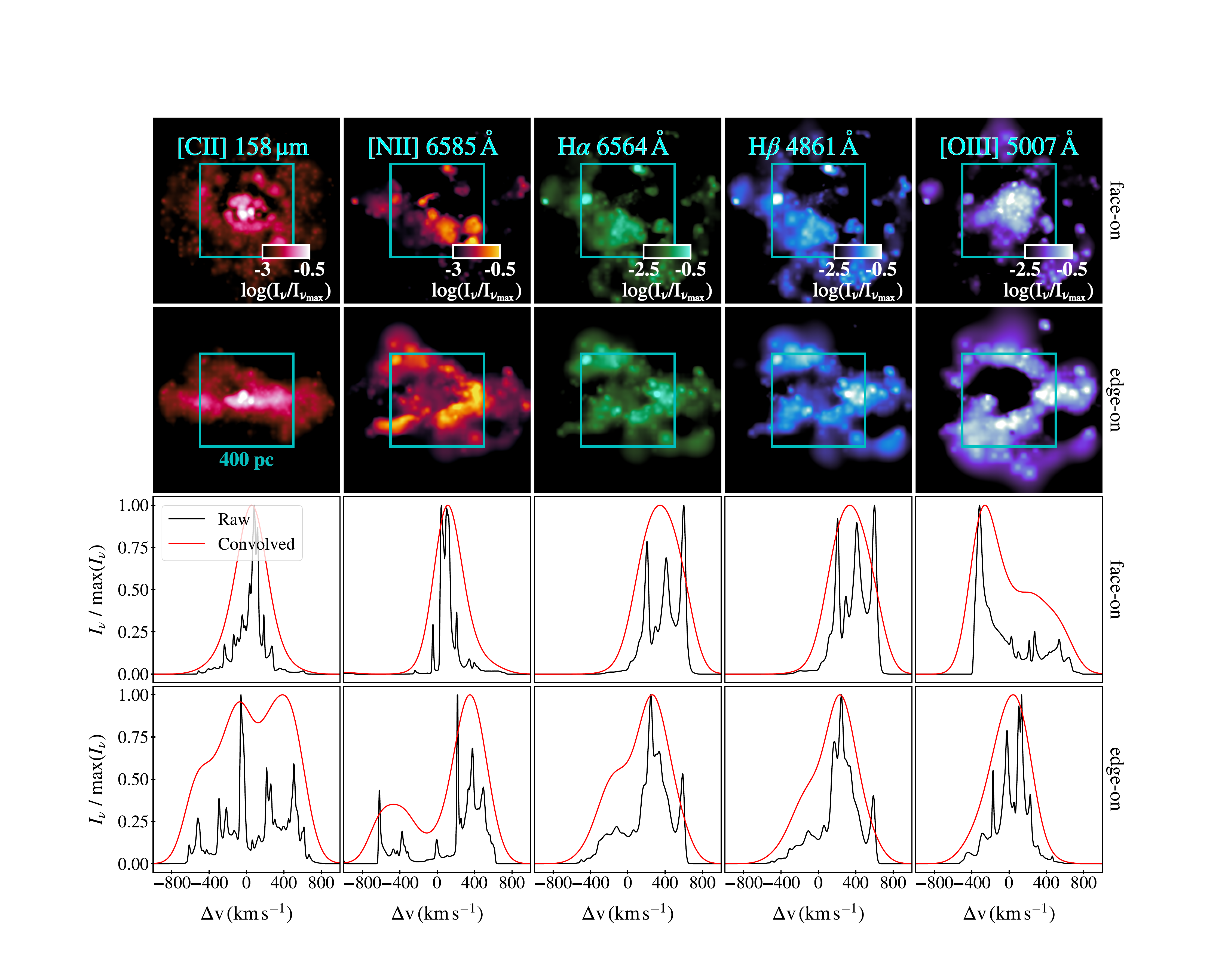}
    \caption{Synthetic IFU observations of galaxy B2 at $z\sim5.4$ for emission lines [\ion{C}{2}]\,158\,$\upmu \rm m$, [\ion{N}{2}]\,6585\,\r{A}, H$\upalpha$, H$\upbeta$, and [\ion{O}{3}]\,5007\,\r{A} (from left to right in order of increasing ionization energy), capturing the morphology and dynamics of the multiphase ISM within the central 400\,pc. The top rows show velocity-integrated images ($\pm$1000\,\kmps~range relative to the stellar center of mass) and the bottom rows show spatially-integrated spectra for the inner 400\,pc region (cyan squares), both for face-on and edge-on projections. Synthetic spectra are shown at full simulation resolution (5\,\kmps{}; black) and degraded to 300\,\kmps~resolution (red). Different emission lines capture rather different morphological and kinematic features, with [\ion{C}{2}]\,158\,\microm{} tracing the colder/denser gas disk and nebular emission lines tracing the hotter/ionized component (significantly impacted by dust attenuation). Bright off-center clumps appear at a wide range of LOS velocities, mimicking broad emission line components at limited (observed) resolution.}
    \label{fig:IFUs}
\end{figure*}

\begin{figure*}
    \includegraphics[width = \textwidth]{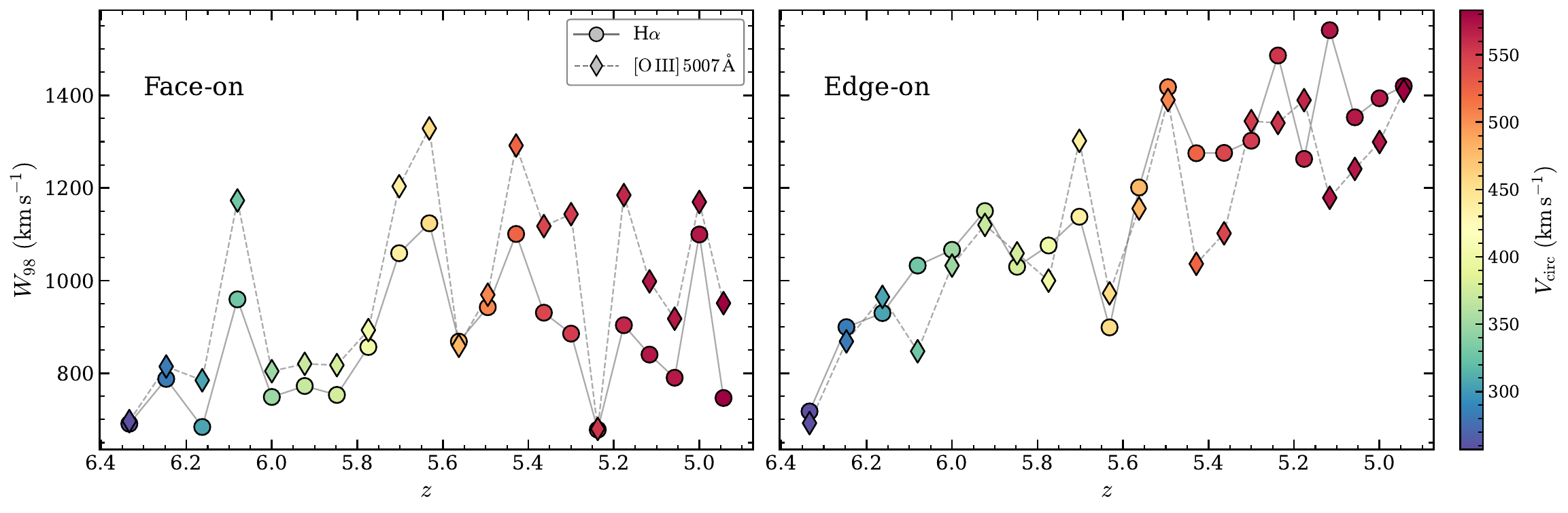}
    \caption{The $W_{98}$ line width of H$\upalpha$ (circles; solid line) and [\ion{O}{3}]\,$5007$\,\r{A} (diamonds; dashed line) as a function of redshift, shown for the face-on (\textit{left}) and edge-on (\textit{right}) projections of galaxy B2 during ultra-compact phases. Line profiles are extracted within an aperture of $400\,\mathrm{pc}$ centered on the host and convolved with a Gaussian line-spread function with $\mathrm{FWHM}=300\,\mathrm{km\,s^{-1}}$; $W_{98}$ is then measured directly from the convolved profile, with no deconvolution applied. The markers are coloured by the host circular velocity $V_{\mathrm{circ}}$ (for $R<100$\,pc). 
    Edge-on line widths follow a near-monotonic trend with redshift, broadening as $V_{\mathrm{circ}}$ increases at lower redshift, whereas face-on line widths tend to be less extreme and exhibit larger redshift-to-redshift scatter. 
    H$\upalpha$ and [\ion{O}{3}]\,$5007$\,\r{A} line widths agree to within the scatter and line-of-sight variation at most epochs. 
    }
    \label{fig:Halpha-OIII-linewidth-evolution}
\end{figure*}

In the right panel of Figure~\ref{fig:SEDs}, we show the Balmer break strength as a function of redshift for all simulated galaxies during their early ultra-compact phases (identified as $R_{\rm eff} < 300\,$pc and $V_{\mathrm{circ}}\,>\,500$\,\kmps). For each selected snapshot, we compute SEDs for seven different LOS and show the median Balmer break strength (solid lines) and the full range of variation (shaded region). For comparison, we also show Balmer break strengths from the corresponding dust-free SEDs (dotted lines).
Encouragingly, we see strong Balmer breaks across all simulated galaxies selected to be as compact as inferred LRD sizes. 
These Balmer break strengths ($\sim$2) are approximately twice as strong as those of typical galaxies ($\sim$1) at similar redshifts, selected from \citet{Roberts-Borsani2024ApJ...976..193R}, and consistent with observed LRDs \citep{wang2024ApJ...969L..13W, setton2025ApJ...995..118S}.
Interestingly, some quiescent galaxies at $z \gtrsim 5$ show Balmer break strengths similar to LRDs \citep[$\sim$1.5--2.5;][]{Strait2023ApJ...949L..23S, deGraaff2025NatAs...9..280D, weibel2025ApJ...983...11W}, while our simulated galaxies are all star-forming during the selected massive and ultra-compact phases that we explore here.

To further characterize the physical conditions and observational signatures of our LRD-like ultra-compact galaxies, Figure~\ref{fig:IFUs} presents synthetic IFU observations of galaxy B2  at $z \sim 5.4$ for the key emission lines [\ion{C}{2}]\,158\,$\upmu \rm m$, [\ion{N}{2}]\,6585\,\r{A}, H$\upalpha$, H$\upbeta$, and [\ion{O}{3}]\,5007\,\r{A}. The upper panels show intensity maps integrating emission lines over the $\pm$1000\,\kmps~velocity range, defined relative to the center of mass of the stellar component.  Emission lines are ordered by increasing ionization potential, from left to right, and we show both face-on and edge-on projections. These velocity-integrated intensity maps reveal a clumpy multiphase ISM where the different gas phases are significantly spatially segregated. The [\ion{C}{2}]\,158\,\microm{} emission, tracing the comparatively cold ($T<10^4\,$K) and dense gas reservoir, is tightly concentrated within the central nuclear disk. In contrast, nebular lines originating from hotter, ionized regions, such as H$\upalpha$, H$\upbeta$, and [\ion{O}{3}]\,5007\,\r{A} exhibit more extended and diffuse morphologies, which is particularly apparent in the edge-on view.

The lower panels of Figure~\ref{fig:IFUs} show spatially-integrated spectra for the inner $400\,$pc, shown at both simulation resolution (5\,\kmps; black solid line) and typical observational resolution ($300\,$\kmps; red solid line) for face-on and edge-on orientations. Total flux is conserved during the convolution to achieve typical observational resolution. The synthetic spectra demonstrate the extreme kinematic state of the galaxy across all gas phases. The raw, high-resolution spectra reveal significant velocity broadening, extending up to $\pm 800\,$\kmps~for edge-orientations, which is consistent with the galaxy's extreme central gravitational potential analyzed in Figure~\ref{fig:sfr-weighted-hist-and-rotation}. Crucially, the high-resolution spectra also display highly complex substructure with multiple distinct peaks in all emission lines, directly tracing numerous discrete, massive, and high-velocity gas clumps that are resolved in the simulation. However, these substructures become merged after convolution to typical IFU resolution, resulting in smoother but still very broad profiles that can reach $\text{FWHM} > 1000\,$\kmps. This suggests that massive, ultra-compact galaxies in the early universe can naturally produce broad Balmer lines similar to observed LRDs through a combination of substantial galaxy-scale broadening from the global mass distribution and the kinematic signatures of multiple discrete, high-velocity gas clumps. However, broad-line AGN are often identified by broad H$\upalpha$ alongside a narrow [\ion{O}{3}]\,5007\r{A}; the widths of these two lines should therefore be systematically compared during ultra-compact phases to test whether this canonical broad-narrow signature can be reproduced serendipitously through galaxy-scale dynamics alone.

\begin{figure*} 
    \centering 
    \includegraphics[width = \textwidth]{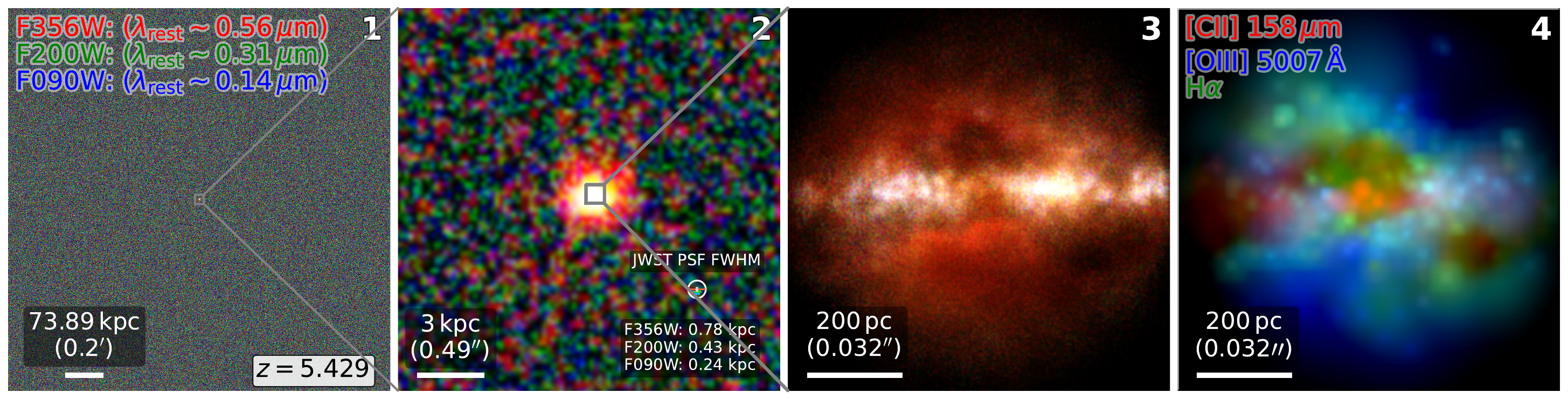}
    \caption{Synthetic RGB images of galaxy B2 at $z\sim 5.4$ from an edge-on LOS, focusing on broadband images in JWST NIRCam broadband filters F356W\,(R), F200W\,(G), and F090W\,(B) on the left three panels and emission line maps in the rightmost panel (showing [\ion{C}{2}]\,158\,\microm\,(R), H$\upalpha$\,(G), and [\ion{O}{3}]\,5007\,\r{A}\,(B)). The broadband RGB images are presented for three different fields of view, gradually zooming in from a full NIRCam field ($2.2^{'}\times2.2^{'}$) to the central 400\,pc region around the ultra-compact galaxy, where the images in panels 1 and 2 include realistic PSFs and noise. 
    When compared to NIRCam PSFs (panel 2), stellar light from the central source would be at most marginally spatially resolved in the rest-frame UV and optical. Panel 3 exhibits a high-resolution, noise-free version of the broadband filter images, where the disk-dominated morphology of the stellar component shows dust obscuration and asymmetric emerging stellar light. In panel 4, we highlight the striking difference in the spatial distribution of multi-phase gas as traced by different emission lines, where [\ion{C}{2}]\,158\,\microm{} shows a more compact morphology tracing colder and denser gas compared to the more extended H$\upalpha$ and [\ion{O}{3}]\,5007\,\r{A} emission.}
    \label{fig:RGBs}
\end{figure*}

Figure~\ref{fig:Halpha-OIII-linewidth-evolution} shows the width of the H$\upalpha$ and [\ion{O}{3}]\,5007\,\r{A} emission lines corresponding to face-on (left) and edge-on (right) projections as a function of redshift in all ultra-compact snapshots identified for galaxy B2. Here we use non-parametric $W_{98}$ wing widths, based on the velocity range encompassing 98\%~of the line flux, to quantify broad emission without decomposing the line into narrow and broad Gaussian components. Along edge-on LOS, $W_{\rm 98}$ rises from $\sim 700\,$km\,s$^{-1}$ at $z \simeq 6.4$ to $\sim 1400\,$km\,s$^{-1}$ by $z \simeq 5.0$ in both tracers, following the growth of $V_{\rm circ}$ (indicated by the marker color) during this redshift interval. In contrast, face-on LOS show significant stochasticity, with line widths fluctuating in the range $W_{\rm 98} \approx 700$--1300\,km\,s$^{-1}$ regardless of the overall monotonic increase of $V_{\rm circ}$. Emission lines are generally broader in edge-on compared to face-on LOS, as expected given the dominant rotational support (Figure~\ref{fig:sfr-weighted-hist-and-rotation}), but the large $W_{\rm 98}$ seen in face-on LOS indicates substantial turbulence and stochastic bulk motions. The line broadening is therefore primarily set by the depth of the gravitational potential and by the orientation of the sightline with respect to the disk, with significant stochasticity introduce by extraplanar gas motions.
The H$\upalpha$ and [\ion{O}{3}]\,5007\,\r{A} lines trace different ionized gas kinematics, but their line widths generally correlate throughout the ultra-compact phase. Interestingly, the $W_{\rm 98}$ of [\ion{O}{3}]\,5007\,\r{A} systematically exceeds that of H$\upalpha$ for face-on LOS, with differences up to $\sim$300\,km\,s$^{-1}$, while H$\upalpha$ is generally broader for edge-on projections. In a few instances, the galaxy exhibits up to $\sim$30\% broader H$\upalpha$ compared to [\ion{O}{3}]\,5007\,\r{A}, but the widths of the two lines remain comparable, suggesting that galaxy scale dynamics alone cannot reproduce the canonical broad H$\upalpha$--narrow [\ion{O}{3}]\,5007\r{A} signature of standard broad-line AGN.

\indent Further exploring the morphological complexity of simulated ultra-compact galaxies, Figure~\ref{fig:RGBs} shows synthetic observations of galaxy B2 at $z \sim 5.4$ for an edge-on LOS in both broadband NIRCam-like images in filters F356W\,(R), F200W\,(G), and FW090\,(B); and narrowband IFU for different emission lines -- [\ion{C}{2}]\,158\,$\upmu \rm m$\,(R), H$\upalpha$\,(G), and [\ion{O}{3}]\,5007\,\r{A}\,(B). The three broadband panels zoom in from the full NIRCam field ($2.2^{'}\times2.2^{'}$) to the inner $400$\,pc of the ultra-compact galaxy. Once we apply the JWST PSF for each of the channels and add 27.5\,mag\,arcsec$^{-2}$ Gaussian noise (panels 1 and 2), the source is clearly detected in all three broadband filters, but it is at most marginally spatially resolved: the ratio of the measured image FWHM to the corresponding PSF FWHM is approximately $2.3$, $1.5$, and $1.3$ in F090W, F200W, and F356W, respectively. The rest-frame optical emission in F356W is therefore only slightly broader than the instrumental PSF, while the rest-frame UV in F090W is the most extended relative to its PSF, reflecting residual structure that survives the beam at the shortest wavelength compared to the more genuinely compact optical core.

Panel 3 exhibits the same broadband filter image without the PSF and noise, and for a small field of view around the center of the galaxy. In this high-resolution image, the galaxy shows a disk-dominated morphology in the stellar component, with the emerging stellar light clearly impacted by dust absorption and scattering in foreground gas clumps, resulting in a very asymmetric light distribution. Interestingly, in addition to the galaxy center, there is prominent dust obscuration of rest-frame UV/optical above and below the plane of the disk, spatially coinciding with the diffuse extended emission from H$\upalpha$ and [\ion{O}{3}]\,5007\,\r{A}, suggesting the propagation of dusty winds in the polar direction despite the dominant inflow component across the galaxy disk (as seen in Figure~\ref{fig:sfr-weighted-hist-and-rotation}). 

Panel 4 highlights the striking difference in the detailed spatial distribution of [\ion{C}{2}]\,158\,$\upmu \rm m$, H$\upalpha$, and [\ion{O}{3}]\,5007\,\r{A} driven by the different gas phases that dominate each emission line. In particular, this demonstrates that H$\upalpha$ and [\ion{O}{3}]\,5007\,\r{A} can probe substantially different gas distributions, with [\ion{O}{3}]\,5007\,\r{A} appearing more extended than H$\upalpha$ in this case, and can therefore show different galaxy-scale kinematics, as seen in Figure~\ref{fig:Halpha-OIII-linewidth-evolution}.

\begin{figure}
    \centering
    \includegraphics[width=1\linewidth]{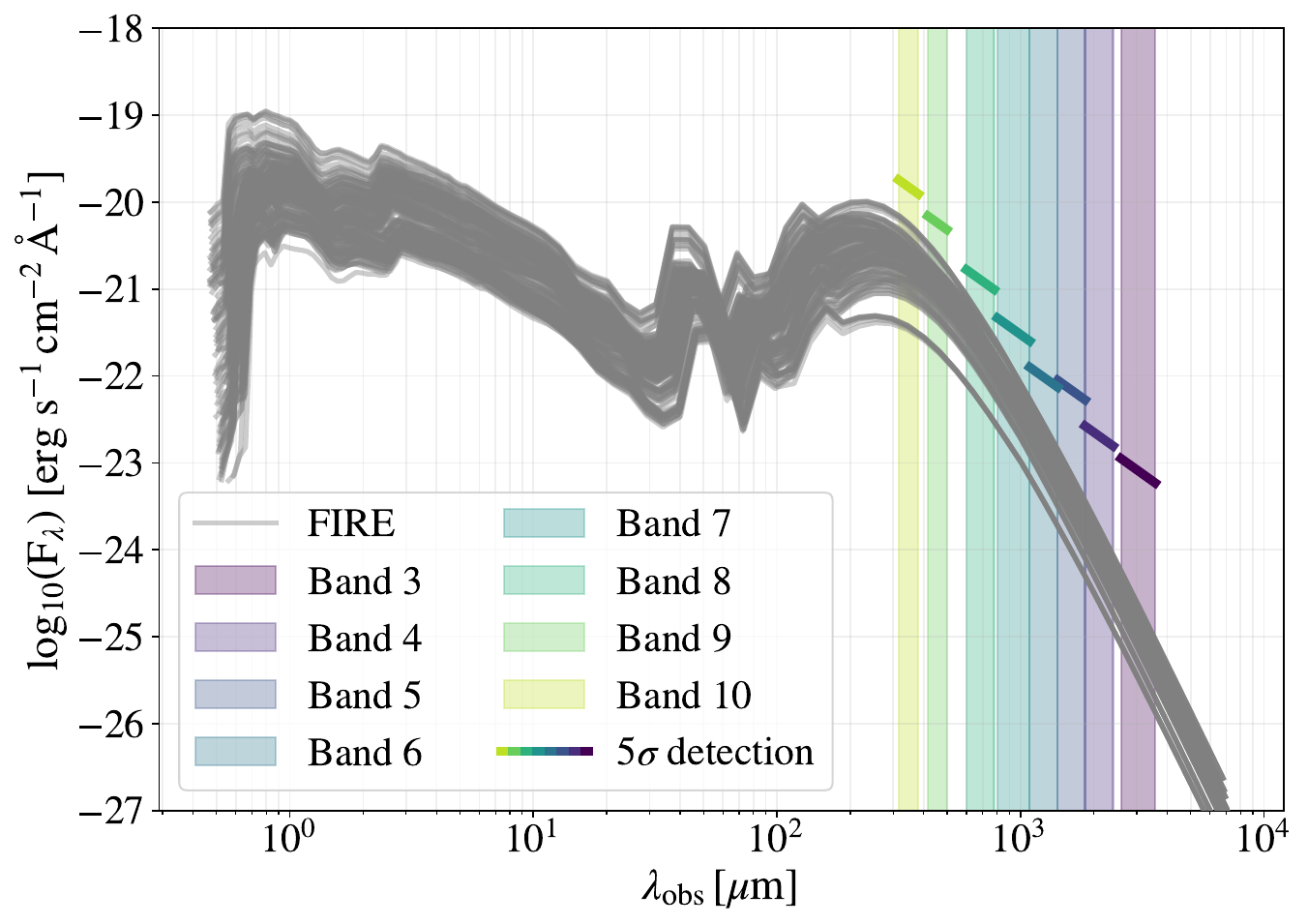}
    \caption{ALMA detectability of dust emission in galaxy B2 during its ultra-compact phase from $z \approx 4.8$--6.4. Grey lines represent SEDs for seven different LOS across the full redshift range. All eight ALMA bands are shown along with their 5$\sigma$ detection limit assuming one-hour-long integration times. This comparison shows that during the simulated ultra-compact phases, the thermal dust emission in the rest frame mid-to-far-IR wavelength range would be undetected by ALMA in reasonable integration times.}
    \label{fig:ALMA}
\end{figure}

Given the clear impact of dust attenuation on our synthetic broadband photometry and spectra, it is important to compare the thermal emission from dust against available constraints. Figure~\ref{fig:ALMA} quantifies the detectability of far-IR dust continuum emission from galaxy B2, comparing SEDs across the ultra-compact phase ($z = 4.8\text{--}6.4$) for seven different inclinations (gray lines) to the sensitivity limits of different ALMA bands. The predicted flux densities in the rest-frame mid-to-far-IR (far-IR to sub-mm in observed frame) regime fall systematically below the $5\sigma$ detection thresholds for one-hour-long integration times in ALMA Bands 3 through 10 (from purple to yellow). Even at the peak of the thermal dust emission, the predicted fluxes are roughly one order of magnitude fainter than the sensitivity limits. 
We find that the mean ALMA integration time required for a $5\sigma$ dust continuum detection across Bands~6--10 during ultra-compact phases is $\sim$150\,hr, $\sim$30\,hr, $\sim$2\,hr, and $\sim$7\,hr for halos B1, B2, C1, and C2, respectively. This suggests that the total dust mass of such high-z ultra-compact galaxies remains insufficient to produce easily-detectable sub-millimeter continuum emission under standard observational constrains of ALMA, but some ultra-compact galaxies (such as C1 in our simulated sample) could be detected without exceptionally deep integration.

\section{Discussion and Conclusions}
\label{sec:discussion-conclusion}

We have shown that ultra-compact galaxy phases at $z>4$ arise naturally in massive, early-forming systems in the \FIREtwo simulations, where sustained gas inflow into the central $\gtrsim$100\,pc and dissipative interactions trigger intense nuclear star formation, rapidly building a compact stellar core with effective radius $R_{\rm eff} < 300$\,pc and stellar mass $M_{\star} \sim 10^{8.5}$--$10^{10.5}$\,M$_{\odot}$,
reaching extreme circular velocities $V_{\rm circ} > 500\,$\kmps. 
The early formation of compact central stellar systems is also seen in other high-redshift cosmological simulations, where gas-rich compaction events driven by mergers, counter-rotating streams, and disk instabilities \citep{Ceverino2015MNRAS.447.3291C,Zolotov2015MNRAS.450.2327Z,Tacchella2016MNRAS.458..242T,Lapiner2023MNRAS.522.4515L,Mayer2024ApJ...961...76M,Andalman2025MNRAS.540.3350A,dekel2025arXiv251107578D,Kohandel2025A&A...704A..39K,Cataldi2026A&A...706A.125C,Liempi2026arXiv260603375L} or the coherent wind-driven reaccretion of gas accumulated in the inner CGM \citep{Mercedes-Feliz2025arXiv251019905M} can rapidly concentrate baryons on $\sim$100\,pc scales and produce dense nuclear stellar components. While their detailed properties depend on subgrid physics, halo mass, and redshift, the qualitative agreement across different numerical methods suggests that compact stellar phases are a generic outcome of rapid, dissipative galaxy assembly at high redshift. Importantly, the ultra-compact galaxy properties presented here emerge without tuning the simulations to reproduce such extreme conditions at early times, implementing identical ISM physics, star formation, and stellar feedback models tested extensively in other \FIREtwo simulations and shown to match a broad range of galaxy observables down to $z=0$ \citep{hopkins2018MNRAS.480..800H}. This represents a crucial robustness test of whether LRD-like stellar systems can arise within a galaxy formation model constrained well beyond the high-redshift regime.

The fact that these ultra-compact galaxy phases broadly reproduce the inferred sizes and stellar masses of observed LRDs \citep{akins2025ApJ...991...37A,Carranza-Escudero2025ApJ...989L..50C,leung2025ApJ...992...26L,rinaldi2025ApJ...992...71R} suggests that at least some LRDs may represent a short-lived and physically expected phase of rapid galaxy assembly at $z\sim4$--8 rather than intrinsically exotic objects. However, establishing whether this is a viable formation channel for LRDs requires confronting the simulations with additional observational constraints, including the full UV-to-optical SED shapes, emission-line widths, dust emission and attenuation, number densities, and host-halo environments. As we discuss below, our combined results suggest that stellar-only scenarios are unable to reproduce some key LRD characteristics and support hybrid models with compact stellar and AGN components.

\subsection{Host halo masses and demographics}

The ultra-compact phases identified here depend strongly on halo growth history, occurring only in the most massive systems in our simulation sample, which are the progenitors of group-size halos with $M_{\rm halo} \gtrsim 10^{13.5}$\,M$_\odot$ at $z=0$ \citep{feldmann2017colours}. Galaxies in lower-mass halos continue to grow, but do not become sufficiently compact early enough to match the inferred LRD sizes. This suggests that LRDs formed through an ultra-compact galaxy channel would preferentially arise in overdense environments, or correspond to the more massive and UV-luminous subset of the observed population. In our simulations, ultra-compact galaxies have UV luminosities $-23\lesssim\,M_{\rm UV}\lesssim\,-20$ (computed as described in Section~\ref{subsec:SED-methods}), placing them near the bright end of the observed LRD UV luminosity function \citep{kokorev2024ApJ...968...38K}. After this transient phase, lasting $\sim$150--400\,Myr, the simulated galaxies grow inside-out and evolve toward more extended massive systems, plausibly including the progenitors of present-day group central galaxies.

During the ultra-compact phases, the host halos of our simulated LRD analogs span $M_{\rm halo} \sim 10^{11}$--$10^{12.5}$\,M$_\odot$ from $z \approx 8 \rightarrow 4$, with typical values $M_{\rm halo} \sim 10^{12}$\,M$_\odot$. These halo masses are consistent with the clustering-based estimate for the LRD in the overdense field at $z=7.3$ analyzed by \citet{Schindler2025NatAs.tmp..191S}, but they are generally higher than the characteristic halo masses inferred for broader LRD samples. Clustering analyses of JWST-selected broad-line AGN overlapping with the LRD population infer $M_{\rm halo} \approx 10^{11.5}$\,M$_\odot$ at $z\sim5$--$6$ \citep{Arita2025MNRAS}, while \citet{Pizzati2025MNRAS} argue that LRDs are too abundant to reside exclusively in quasar-like halos and consider viable host thresholds as low as $M_{\rm halo} \sim 10^{11}$\,M$_\odot$, depending on duty cycle. Similarly, \citet{Carranza-Escudero2025ApJ...989L..50C} find that LRDs tend to occupy lower-density environments than galaxies of similar redshift and infer low characteristic halo masses under their clustering interpretation. These results suggest that the broader LRD population extends to lower halo masses and less biased environments than those sampled by our ultra-compact analogs.

However, abundance estimates suggest that this channel is not obviously too rare to contribute to the observed abundance of bright LRDs, although the estimate remains highly uncertain. Halos with present-day masses $M_{\rm halo}(z=0)>10^{13.5}$\,M$_\odot$ have a comoving number density $\sim 10^{-4}$\,cMpc$^{-3}$. If their main progenitors undergo an LRD-like ultra-compact phase with a duty cycle of $\sim$20\%, motivated by an active duration of $\sim$200\,Myr over the $\sim$1\,Gyr interval from $z\sim8 \rightarrow 4$, the implied number density is $\sim 2 \times 10^{-5}$\,cMpc$^{-3}$. This is comparable to published LRD abundance estimates $\sim 10^{-5}$\,cMpc$^{-3}$ in the redshift range $z \sim 3$--8 \citep{greene2024ApJ...964...39G,kokorev2024ApJ...968...38K,Pizzati2025MNRAS} but below the $\sim 10^{-4}$\,cMpc$^{-3}$ peak reported at $z\sim5$--6 by \citet{Carranza-Escudero2025ApJ...989L..50C}. Overall, the abundance of massive halo progenitors would be sufficient to explain a non-negligible fraction of the bright LRD population, at least at the order-of-magnitude level, but the full LRD population likely extends to lower halo masses than our simulated analogs.

AGN LRD scenarios do not require such massive host halos, but they do require a sufficiently abundant population of actively accreting BHs in early low-mass galaxies. This is not guaranteed theoretically, given current discrepancies between cosmological simulations \citep{Habouzit2022MNRAS.511.3751H,Habouzit2025MNRAS.537.2323H,Izquierdo-Villalba2026} and physical effects such as bursty stellar feedback repeatedly evacuating nuclear gas reservoirs and suppressing early BH growth \citep{dubois2015MNRAS.452.1502D, habouzit2017MNRAS.468.3935H, angles-alcazar2017MNRAS.472L.109A, byrne2023MNRAS.520..722B}. Using FIRE-2 simulations spanning a broader range of host galaxy masses than considered here, \citet{Marszewski_2026} showed that this is not a limiting factor under efficient BH accretion scenarios, where low-luminosity AGN are indeed over-predicted and the LRD bolometric \citep{greene2026ApJ...996..129G} and UV \citep{kokorev2024ApJ...968...38K} luminosity functions can be reproduced by restricting LRDs to the subset of AGN with super-Eddington accreting black holes \citep[e.g.,][]{madau2024ApJ...976L..24M, Pacucci_2024, quadri2025A&A...704A.248Q, zhang2025ApJ...995...26Z, chen2026arXiv260531077C} residing in galaxies with $M_\star \gtrsim 2 \times 10^7$\,M$_\odot$. AGN-dominated scenarios can thus more easily accommodate a population of LRDs in lower mass systems, but demographics alone is insufficient to rule out the potential contribution of ultra-compact galaxies at the UV-bright/massive end of the LRD distribution.

\subsection{SEDs, Balmer breaks, and UV/optical slopes}

Our full 3D dust-continuum radiative transfer calculations show that strong Balmer breaks do not uniquely imply either quiescence or a gas-enshrouded AGN continuum. Interestingly, our simulated galaxies consistently show Balmer break strengths $\sim$2 during the ultra-compact phases, overlapping with observed LRDs \citep{wang2024ApJ...969L..13W, setton2025ApJ...995..118S} and some high-redshift quiescent galaxies \citep{Strait2023ApJ...949L..23S, deGraaff2025NatAs...9..280D, weibel2025ApJ...983...11W}. While simulated galaxies remain actively star-forming during the ultra-compact phases, a substantial component of intermediate-age stars with ages $\sim$100\,Myr--1\,Gyr  (Figure~\ref{fig:stellar-age-bins}) naturally creates Balmer break strengths comparable to those inferred for LRDs, except for extreme cases above the maximum limit ($\sim$3) that stellar populations alone cannot produce \citep{naidu2025arXiv250316596N, deGraaff2025A&A...701A.168D, rusakov2026Natur.649..574R}. Importantly, the synthetic SEDs reproduce the blue UV slopes of observed LRDs, suggesting that the negative $\beta_{\rm UV}$ can be largely stellar in origin rather than arise from scattered AGN light, consistent with recent arguments that the UV excess in LRDs is difficult to explain as pure scattered AGN emission \citep{bao2025ApJ...992..117B} and favoring a single-component evolved stellar population dominating the rest frame UV-optical range in LRDs along with the recurrent inflection point at the Balmer limit \citep{setton2025ApJ...995..118S}. This result has important implications for stellar mass estimates: if a substantial fraction of the rest-frame UV continuum is stellar, then AGN-dominated models may underestimate the stellar content of LRDs. 
Interestingly, including an AGN component in SED modeling does not always lower the inferred stellar mass. Some studies find stellar masses comparable to those from stellar-only models \citep{Carranza-Escudero2025ApJ...989L..50C} or even higher if adding AGN requires a larger dust attenuation \citep{rinaldi2025ApJ...992...71R}, consistent with the stellar masses of our ultra-compact galaxies.

On the other hand, our ultra-compact LRD analogs fail to reproduce the positive rest-frame optical slope that completes the characteristic V-shaped continuum observed in LRDs \citep{Killi2024A&A...691A..52K, Wang2025ApJ...984..121W}, requiring an additional spectral component.  Providing red optical emission without washing out the common inflection point near the Balmer break is difficult to achieve with a simple unobscured broken power-law AGN dominating the entire UV-to-optical SED \citep{setton2025ApJ...995..118S}. A more plausible interpretation is thus a hybrid model where stars dominate the UV continuum and the Balmer break, while an obscured or reprocessed AGN contributes to the redder optical continuum, as supported by other recent models based on cosmological hydrodynamic simulations \citep{Volonteri2025A&A...695A..33V,quadri2025A&A...704A.248Q,lachance2026OJAp....955493L,Marszewski_2026}. In this scenario, the AGN component will more likely be emission from an SMBH enshrouded by a gas cocoon or a super-Eddington accreting BH rather than a standard template \citep{greene2026ApJ...996..129G}. Recent morphological analysis by \citet{cloonan2026arXiv260324700C} has shown that the rest-UV light in LRDs is typically more extended (though still compact; $R_{\rm eff} \sim 200$\,pc) than the rest-optical ($R_{\rm eff} \lesssim 100$\,pc), further supporting a stellar origin of UV light from ultra-compact hosts even if an unresolved AGN dominates the rest-optical regime.

\subsection{Dust mass, attenuation, and far-IR constraints}

Adopting a dust treatment with fixed dust-to-metals ratio and a Milky Way grain model \citep{cochrane2024ApJ...961...37C}, our synthetic SEDs show substantial dust attenuation, shaping the UV--optical continuum, decreasing UV luminosities, and enhancing the apparent Balmer break strength.
The predicted suppression of intrinsic UV luminosity by a factor $\sim$100 in our simulated analogs is broadly consistent with SED-based interpretations of LRDs inferring high stellar masses and significant dust attenuation \citep[e.g.,][]{labbe2023Natur.616..266L, akins2025ApJ...991...37A, rinaldi2025ApJ...992...71R}.
Given the impact of dust on synthetic observables, it is crucial to understand the nature and amount of dust in LRDs through multi-wavelength approaches from rest-frame UV/optical to Mid-IR and Far-IR. The total dust mass, grain composition, and geometry of LRDs remain uncertain, but inferred average dust masses $\sim10^4\,\rm M_{\odot}$, with upper limit $\sim10^6\,\rm M_{\odot}$ regardless of AGN/stellar modeling \citep{Casey2024ApJ...975L...4C, casey2025ApJ...990L..61C}, suggest low dust budgets in LRDs \citep[see also][]{chen2025ApJ...994L..42C}.  Under our simplified dust model ($M_{\rm dust}=0.4\times M_{\rm metals}$), simulated ultra-compact galaxy phases have average dust masses
$M_{\rm dust}\sim 10^6\,\rm M_{\odot}$, consistent with observed upper limits. In addition, forward-modeled ALMA fluxes fall below the 5$\sigma$ detection threshold for one-hour-long exposures, consistent with current upper limits and non-detections of Far-IR emission in LRDs with deep ALMA observations \citep{casey2025ApJ...990L..61C}.

\subsection{Emission line properties}

Simulated ultra-compact galaxies can also reproduce some emission line properties of observed LRDs.  Our kinematic analysis  and full 3D line radiative transfer calculations show that broad Balmer lines are not, by themselves, definitive evidence for a classical BLR.  During the ultra-compact phases, the deepening of the central gravitational potential, coherent rotation, strong turbulence, radial inflow, and clumpy gas distributions combine to produce Balmer line widths up to  $W_{98} \sim 1500$\,\kmps~after convolution to JWST-like spectral resolution. In this regime, unresolved substructure can blend into a smooth broad component, suggesting that part of the line width often attributed to virialized gas in the BLR around the SMBH may instead reflect galaxy-scale gas dynamics, as suggested by \citet{baggen2024ApJ...977L..13B}. This may have important consequences for BH mass estimates: if some fraction of the observed Balmer line width is set by host-galaxy kinematics rather than a sub-pc BLR, then single-epoch virial SMBH masses could be biased high in at least part of the LRD population. 

However, Balmer line broadening arising purely from galaxy kinematics is unlikely to produce very symmetric line profiles, both in blue and red wings, as observed in many LRD sources. This symmetry, therefore, serves as a key diagnostic: broadening driven by SMBH accretion or electron scattering naturally yields symmetric profiles, whereas a purely kinematic origin would manifest as asymmetric emission from discrete gaseous clumps, distinguishable at sufficient spectral resolution.
Reproducing the full observed range of Balmer line widths under the ultra-compact galaxy scenario would also require adding an AGN component, since emission lines with FWHM $\gtrsim 2000$ km s$^{-1}$ become more difficult to explain based on galaxy-scale dynamics alone. Moreover, standard broad-line AGN are often identified by broad H$\upalpha$ together with comparatively narrow forbidden emission. In our ultra-compact phases, H$\upalpha$ and [\ion{O}{3}]\,5007\,\r{A} generally show comparable non-parametric wing widths, implying that galaxy-scale dynamics broaden both tracers rather than producing a clean broad-Balmer/narrow-forbidden signature. Thus, while ultra-compact galaxy kinematics may contribute to intermediate Balmer widths, the broader and more symmetric Balmer lines seen in many LRDs are more naturally explained by a BLR, a quasi-star phase, or another compact AGN-related configuration, especially when observed along with narrow forbidden lines.

\subsection{Testable predictions}

The multi-phase line transfer and synthetic broad-band imaging yield several observationally testable predictions. In our simulations, [\ion{C}{2}]\,158\,\microm{} is concentrated in the dense neutral nuclear disk while H$\upalpha$, H$\upbeta$, [\ion{N}{2}]\,6585\,\r{A}, and [\ion{O}{3}]\,5007\,\r{A} arise from more extended and dust-attenuated ionized gas. As a result, different emission lines are predicted to trace different morphologies and velocity fields, and some broad components may reflect the superposition of multiple off-center clumps rather than a single central emitting region. A galaxy-driven interpretation of broad lines therefore predicts a complex, spatially varying kinematics in both Balmer and forbidden lines when observed at sufficiently high spatial and spectral resolution, rather than a strict dichotomy between broad permitted lines and narrow forbidden lines. In addition, asymmetric morphologies in broad-band images, strong dust obscuration, and extended ionized structures may be common during the ultra-compact phases. We therefore propose the joint use of [\ion{C}{2}]\,158\microm/H$\alpha$/[\ion{O}{3}]\,5007\,\r{A} spatial mapping, Balmer versus forbidden-line wing widths, and high-resolution line-profile  substructure as a route to uncovering the early progenitors of present-day group-scale systems, with LRD-like sources representing one possible outcome of the selection rather than the only target of interest.

\subsection{Black hole growth and hybrid scenarios}
\label{subsec: black-hole-growth}
While one of the primary goals of this paper is to test the limits of stellar-only LRD scenarios, the physical conditions reached in our ultra-compact galaxies suggest that BHs should indeed become important in these plausible LRD analogs. With typical stellar masses $M_{\star} \sim 10^{8.5}$--$10^{10.5}$\,M$_{\odot}$ within $\sim$300\,pc and a well-defined center, these ultra-compact galaxies are predicted to have formed and retained a central SMBH in most BH seeding and BH dynamics scenarios \citep{ma2021MNRAS.508.1973M, Bhowmick2025ApJ...991...81B, Bonoli2025arXiv250912325B, Bhowmick2026arXiv260612851B}.
The central stellar surface densities approach the empirical and theoretical upper limit for dense stellar systems \citep[$\mathit{\Sigma_\star}\approx3\times10^5$\,M$_{\odot}\rm pc^{-2}$;][]{hopkins2010MNRAS.401L..19H, grudic2019MNRAS.483.5548G}, comparable to nuclear star clusters and ultra-compact dwarfs in the local universe \citep{norris2014MNRAS.443.1151N}, and some models suggest such environments could form BH seeds through stellar collisions even if a BH seed was not already in place \citep{escala2021ApJ...908...57E, 
gonzalez2024ApJ...969...29G,
escala2025ApJ...995...44E, 
bellovary2025ApJ...984L..55B,  
rantala2025MNRAS.542L..78R,
Williams2026arXiv260326872W}. In addition, the values of $M_{\star}$ and $\mathit\Sigma_\star$ reached within the central $\sim$300\,pc imply that these ultra-compact galaxies are in the regime where rapid SMBH growth is expected, with the deep stellar gravitational potential retaining the nuclear gas reservoir against ejection by stellar feedback \citep{dubois2015MNRAS.452.1502D,
habouzit2017MNRAS.468.3935H, 
angles-alcazar2017MNRAS.472L.109A,
grudic2018MNRAS.475.3511G,
grudic2020MNRAS.496L.127G,
byrne2023MNRAS.520..722B,
Sunseri2025arXiv251019822S} and providing strong gravitational torques for efficient inflow down to sub-pc scales \citep{hopkins2010MNRAS.407.1529H,
angles-alcazar2021ApJ...917...53A,
hopkins2024OJAp....7E..18H}.

This suggests that the ultra-compact starburst phase identified here could be a precursor of AGN signatures in LRDs rather than an alternative to early BH growth \citep[see also][]{Liempi2026arXiv260603375L}.  In a hybrid LRD scenario, the ultra-compact stellar system would first generate several key LRD properties, including the strong Balmer break, the blue UV continuum, and moderate broad line widths, while an emerging AGN fueled by the same ultra-compact galaxy conditions would then contribute the red optical continuum and potentially the most extreme Balmer-line broadening. In this picture, the stellar and AGN scenarios are thus complementary rather than mutually exclusive.
This conclusion aligns with the demographics scenario explored by \citet{Marszewski_2026}, where BHs play an essential role in powering the red continuum of LRDs but stellar emission from the host galaxies remains important, especially in the rest-frame UV. The ultra-compact systems identified here may therefore represent the more massive, higher-luminosity population in this hybrid scenario, while the majority of LRDs could still reside in lower-mass galaxies.

\subsection{Possible connection to Little Blue Dots}

The same ultra-compact phases may also be relevant to the emerging population of Little Blue Dots (LBDs; \citealt{Brazzini2026arXiv260122214B, Madau2026arXiv260222386M, Scholtz2026arXiv260322277S, Sneppen2026arXiv260612509S}). Unlike LRDs, LBDs are selected to have blue rest-frame UV ($\beta_{\rm UV} < -0.37$) and optical slopes ($\beta_{\rm opt} < 0$), while still being compact and often showing broad Balmer emission, weak X-ray emission, and little variability. Before adding any AGN component, our simulated ultra-compact galaxies naturally have blue UV and optical slopes, compact sizes, and intermediate Balmer-line widths from galaxy-scale dynamics, suggesting that LBDs could represent an earlier precursor stage or a less AGN-dominated manifestation of the same compact galaxy phase, with subsequent BH growth or reprocessed AGN emission shifting some systems into the LRD color selection by adding a red optical continuum. This interpretation is further supported by auroral-line diagnostics placing LBDs in a region consistent with star-forming sources \citep{Brazzini2026arXiv260122214B}, in line with the ultra-compact starburst channel explored here. However, this plausible evolutionary connection between LBDs and LRDs remains uncertain, as LBDs are often interpreted as type-I AGN candidates with broad Balmer emission and comparatively narrow forbidden lines, while our simulations generally produce comparable H$\upalpha$ and [\ion{O}{3}]\,5007\,\r{A} line widths.

\subsection{Caveats}

The simulations analyzed here do not include AGN feedback, which may thus affect some of the predicted properties of LRD analogs.  The onset of ultra-compact phases is likely robust relative to assumptions in AGN feedback, since central BHs are not expected to accrete efficiently at earlier times, with lower central surface densities driven by stellar feedback and the lack of a well-defined galaxy center.  However, once the galaxy becomes ultra-compact, the duration of this phase and the peak central gas and stellar densities are likely upper limits. The subsequent structural evolution of massive galaxies is very sensitive to feedback, becoming overly compact at lower redshift ($z \lesssim2$) unless strong AGN feedback can regulate the central star formation \citep{dubois2016MNRAS.463.3948D,
choi2018ApJ...866...91C,
wellons2020MNRAS.497.4051W,
parsotan2021MNRAS.501.1591P,
cochrane2023MNRAS.523.2409C,
mercedes-feliz2023MNRAS.524.3446M,
mercedes-feliz2024MNRAS.530.2795M,
byrne2024ApJ...973..149B}, and some degree of AGN feedback could also shorten or perturb the ultra-compact phase at high redshift. The limited sample of simulated halos is another important caveat, which prevents us from predicting the full abundance or demographic mix of LRDs under the hybrid stellar+AGN scenario. While the present study is intended to establish a viable physical channel for the formation of ultra-compact galaxies and evaluate to what extent they can reproduce LRD-like properties, future work should increase the simulated sample to perform statistical analyses.

\subsection{Conclusions}

In summary, we have shown that \FIRE{} cosmological simulations naturally produce massive, ultra-compact, dusty star-forming galaxies that reproduce several of the most striking inferred properties of LRDs, including compact sizes, high stellar masses, strong Balmer breaks, blue UV slopes, lack of sub-millimeter detections, and Balmer-line widths $\gtrsim 1000$\,km\,s$^{-1}$. However, the same models do not reproduce the characteristic red optical continuum and are unlikely to explain more extreme Balmer line widths ($\gtrsim 2000$\,km s$^{-1}$) along with narrow [\ion{O}{3}]\,5007\,\r{A} emission. The transient nature of the ultra-compact phase, its dependence on very early halo growth histories, and their UV luminosities at the upper end of the LRD distribution further imply that an ultra-compact galaxy channel of LRDs would be unlikely to explain the full population. 
These massive, ultra-compact early galaxies are, in fact, predicted to efficiently fuel central BHs, further suggesting that stellar-only scenarios are unable to reproduce the LRD characteristics from cosmological initial conditions.  
These results support a hybrid stellar+AGN scenario, where the LRD spectral features are a combination of galaxy and SMBH, and a heterogeneous interpretation of the LRD population, emphasizing degeneracies in LRD observables that can be reproduced through different physical processes. Independent of the LRD interpretation, the morphological and kinematic diagnostics presented here outline a concrete observational strategy for identifying the ultra-compact progenitors of group-scale halos during their formative $z \sim 4$--$8$ assembly.

\section*{Acknowledgments}
We thank Laura Sommovigo and Rachel Somerville for useful discussions. 
NCR acknowledges support from ACCESS allocation PHY260171. 
DAA acknowledges support from NSF CAREER award AST-2442788, NASA grant ATP23-0156, STScI grants JWST-GO-01712.009-A, JWST-AR-04357.001-A, and JWST-AR-05366.005-A, an Alfred P. Sloan Research Fellowship, and Cottrell Scholar Award CS-CSA-2023-028 by the Research Corporation for Science Advancement. 
RKC is grateful for support from the Leverhulme Trust via the Leverhulme Early Career Fellowship. 
CAFG was supported by NSF through grants AST-2108230 and AST-2307327; by NASA through grants 80NSSC22k0809, 80NSSC22K1124 and 80NSSC24K1224; by STScI through grant JWST-AR-03252.001-A; and by BSF through grant \#2024262.
The FIRE-2 public data repository is hosted at the Flatiron Institute, which is supported by the Simons Foundation.

\bibliography{cite}

\appendix
\label{appendix}

\section{Halos with LRD-like characteristics}\label{sec:allhalos}

\begin{figure}[h!]
    \centering
    \includegraphics[width=\textwidth]{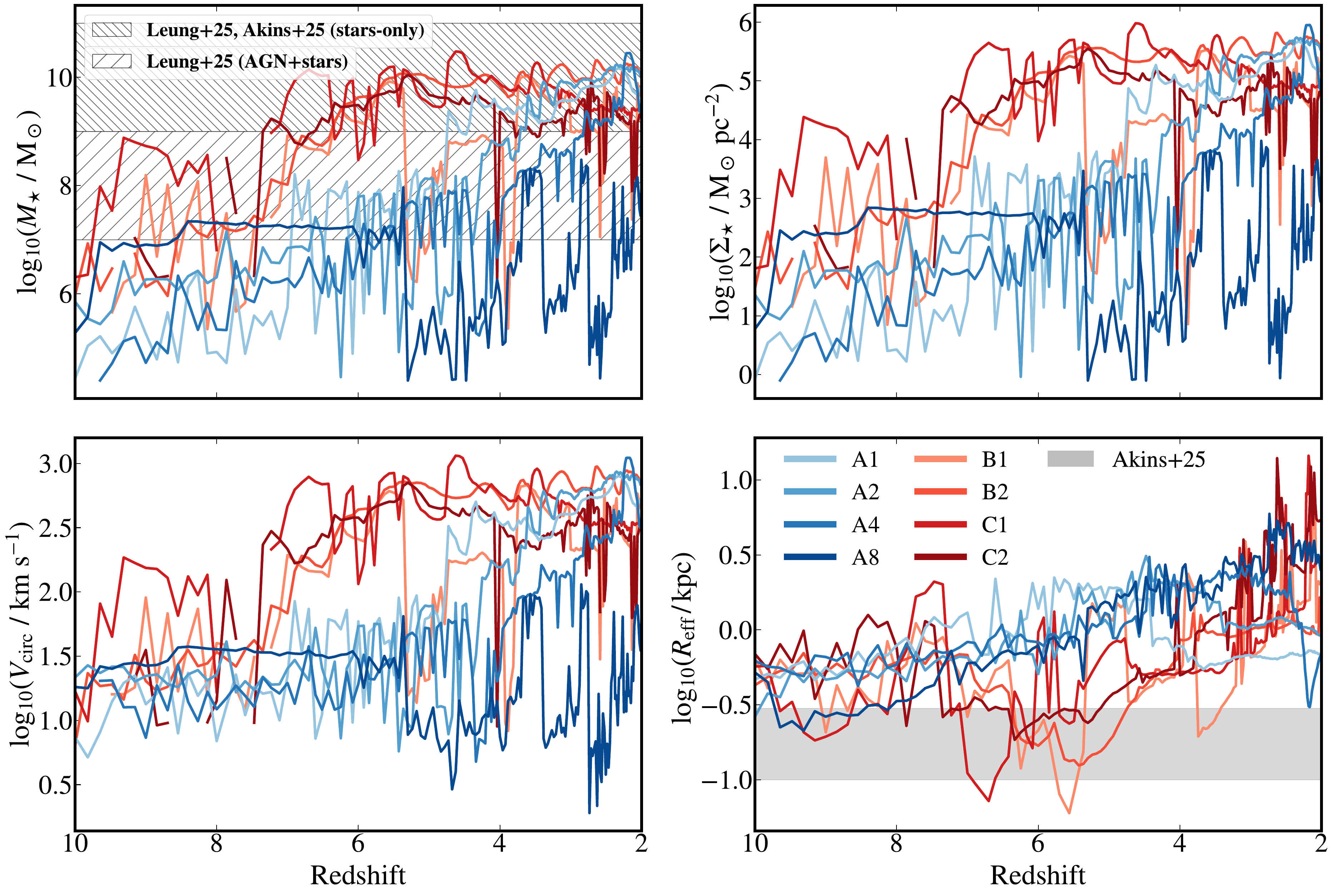}
    \caption{Stellar mass ($M_\star$; top left), stellar mass surface density ($\mathit\Sigma_{\star}$; top right), circular velocity ($V_{\mathrm{circ}}$; bottom left), and stellar half-mass radius ($R_{\mathrm{eff}}$; bottom right) as a function of redshift for all simulated galaxies. All quantities (except $R_{\mathrm{eff}}$) are computed within the central 100\,pc. Observational constraints on the stellar masses and sizes of LRDs are indicated as in Figure~\ref{fig:redshift-evolution}. Only the more massive, earlier growing halos (B1, B2, C1, C2) exhibit a compact stellar population ($R_{\mathrm{eff}}$ $<$ 300\,pc) within the stellar mass and redshift range of observed LRDs (under the stellar-only scenario). B2 is analyzed in more detail throughout the paper as a representative example.}
    \label{fig:all-massive-halos}
\end{figure}

The redshift evolution of the simulated galaxy sample, as shown in Figure  \ref{fig:all-massive-halos}, demonstrates a clear bifurcation in the assembly history of the central regions ($r < 100\,\text{pc}$) based on halo mass. The B- and C-series galaxies (redder colors), which inhabit the most massive progenitors ($M_{\rm halo}\gtrsim10^{13}$\,M$_\odot$ at $z=2$), undergo an accelerated growth phase between $z \approx 4$--8. During this epoch, their central stellar mass ($M_{\star}$; top left) and stellar mass surface density ($\mathit\Sigma_{\star}$; top right) increase by nearly three orders of magnitude while experiencing ultra-compact phases with stellar half-mass radius $R_{\rm eff} < 300$\,pc (bottom right), effectively entering the parameter space occupied by observed LRDs (hatched/shaded regions; \citealt{leung2025ApJ...992...26L, akins2025ApJ...991...37A}). In contrast, the lower-mass A-series halos (bluer colors) exhibit a more stochastic growth history, failing to reach the required $M_{\star}$ and $R_{\rm eff}$ to match observational constraints of LRDs during the early redshifts. The dynamical evolution of galaxies mirrors their mass concentration, as central circular velocities (bottom left) in the more massive halos (B1, B2, C1, C2) frequently exceed $V_{\text{circ}} > 500\,\text{\kmps}$ at $z>4$, reflecting the formation of deep gravitational potential wells that are absent in the lower-mass A-series trajectories.
The episodes of extreme compactness ($R_{\rm eff} < 300$\,pc) are transient in nature, typically persisting for $\sim 150\text{--}400\text{ Myr}$ before galaxies grow inside-out and evolve toward larger radii.

\section{Stellar age distribution}

\begin{figure}[h!]
    \centering
    \includegraphics[width=0.6\textwidth]{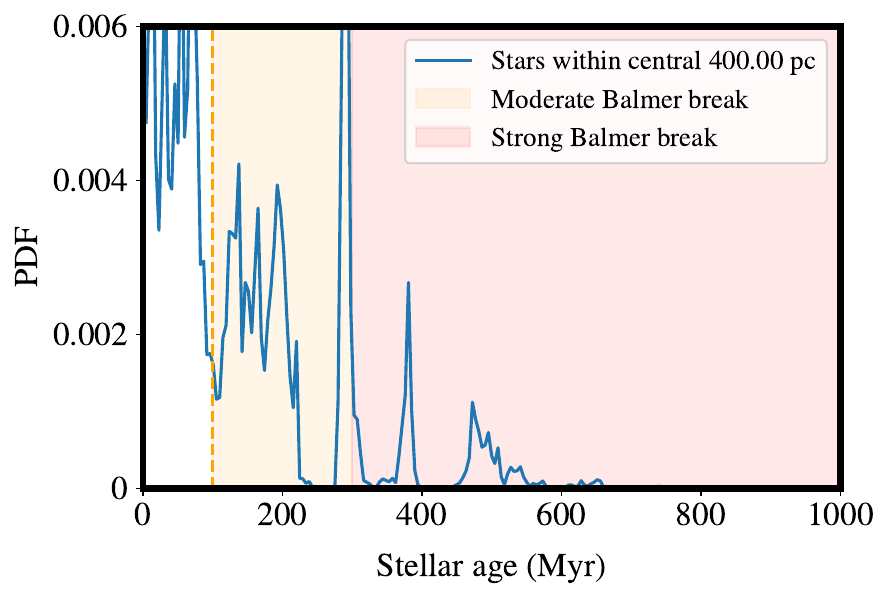}
    \caption{Stellar age distribution for galaxy B2 at $z\sim5.4$ within the central 400\,pc.
    The abundance of stars with ages 100--650\,Myr creates strong Balmer break features in the SEDs observed from different LOS. }
    \label{fig:stellar-age-bins}
\end{figure}

The distribution of stellar ages within the central $400\,\text{pc}$ of galaxy B2 at $z \sim 5.4$ provides a crucial physical link between the galaxy's assembly history and its observed SED. As illustrated in Figure~\ref{fig:stellar-age-bins}, the stellar population exhibits a complex, multi-modal distribution. While there is a significant contribution from young stars ($\lesssim 100\,\text{Myr}$), a substantial mass fraction is concentrated in an intermediate-age population spanning the $100\,\text{Myr}$ to $650\,\text{Myr}$ range. The abundance of stars in the $100\,\text{Myr}$ to $1\,\text{Gyr}$ window falls directly within the range of stellar ages required to produce moderate and strong Balmer break features (shaded yellow and red, respectively; \citealt{poggianti1997A&A...325.1025P}), consistent with the spectroscopic signatures of observed LRDs. 

\end{document}